\documentclass[aps,twocolumn,prd,preprintnumbers, groupedaddress, nofootinbib, 10pt]{revtex4-1}
\usepackage[hidelinks]{hyperref}
\usepackage{amssymb}
\usepackage{amsfonts}
\usepackage{graphicx}
\usepackage{epstopdf}
\usepackage{dcolumn}
\usepackage{amsmath}
\usepackage{latexsym,bm}
\usepackage{amsthm}
\usepackage{slashed}
\usepackage{float}
\usepackage{color}
\usepackage{url}
\usepackage{longtable}
\usepackage{rotating}
\usepackage{ulem}
\usepackage{faktor}

\newcommand{\be}{\begin{equation}}
\newcommand{\ee}{\end{equation}}
\newcommand{\ba}{\begin{aligned}}
\newcommand{\ea}{\end{aligned}}

\newcommand{\bea}{\begin{eqnarray}}
\newcommand{\eea}{\end{eqnarray}}

\def\unit{{1\kern-.65ex {\rm l}}}
\def\1{{1\kern-.65ex {\rm l}}}

\def\bbZ{{\mathbb{Z}}}

\usepackage{verbatim}
\usepackage{tikz}
\usetikzlibrary{arrows,snakes,shapes.arrows,decorations.markings}
     \tikzset{>=triangle 90}
     \tikzstyle{gr}=[draw,circle,green!50!black,fill=green!50!black,scale=.6]
     \tikzstyle{Bl}=[draw,circle,blue,scale=.7]
     \tikzstyle{R}=[draw,circle,fill=red,scale=.7]
     \tikzstyle{bl}=[draw,circle,fill=black,scale=.2]
     \tikzstyle{bbc}=[draw,circle,fill=black,scale=.75]
     \tikzstyle{bbcs}=[draw,circle,fill=black,scale=.5]
     \tikzstyle{rc}=[circle,fill=red,scale=.6]
     \tikzstyle{wc}=[draw,circle,scale=.75]
\usepackage{array} 
\usepackage{multirow} 
\usepackage{colortbl} 
\usepackage{stfloats}  
\usepackage{cellspace}
\usepackage{xcolor}   
\def\blue#1{{\color{blue}{#1}}}
\def\green#1{{\color{black!25!green}{#1}}}

\def\csc{\scriptstyle}

\newcommand{\xdasharrow}[2][->]{
\tikz[baseline=-\the\dimexpr\fontdimen22\textfont2\relax]{
\node[anchor=south,font=\scriptsize, inner ysep=1.5pt,outer xsep=2.2pt](x){#2};
\draw[shorten <=3.4pt,shorten >=3.4pt,dashed,#1](x.south west)--(x.south east);
}
}

\def\^{\wedge}

\def\Z{\mathbb{Z}}

\newcount\hour \newcount\minute
\hour=\time \divide \hour by 60
\minute=\time
\def\now{%
\ifnum \hour<13
  \ifnum \hour=0 \advance \hour by 12 \number\hour:\else \number\hour:\fi%
     \ifnum \minute<10 0\fi%
     \number\minute%
\ A.M.%
\else \advance \hour by -12 \number\hour:%
  \ifnum \minute<10 0\fi%
  \number\minute%
  \ P.M.%
\fi%
}

\makeatother

\begin{document}

\title{4d $\mathcal{N}=2$ S-folds}

\author{Fabio Apruzzi}
\author{Simone Giacomelli}
\author{Sakura Sch\"afer-Nameki}

\affiliation{Mathematical Institute, University
of Oxford, Woodstock Road, Oxford, OX2 6GG, United Kingdom}


\begin{abstract}
\noindent We propose a generalization of S-folds to 4d $\mathcal{N}=2$ theories. 
This construction is motivated by the classification of rank one 4d $\mathcal{N}=2$ super-conformal field theories (SCFTs), which we reproduce from D3-branes probing a configuration of $\mathcal{N}=2$ S-folds combined with 7-branes. 
The main advantage of this point of view is that realizes both Coulomb and Higgs branch flows and allows for a straight forward generalization to higher rank theories. 

\end{abstract}



\maketitle


\section{Introduction}
\label{sec:intro}

The classification of superconformal field theories (SCFTs) with 8 supercharges has seen immense advances in the past years. 
Starting with the chiral supersymmetric theories in 6d, where a classification framework has been developed in F-theory on elliptic Calabi-Yau threefolds \citep{Heckman:2013pva, Heckman:2015bfa, Bhardwaj:2015xxa}, recent years have seen a concerted effort  to establish a classification of 5d $\mathcal{N}=1$ SCFTs constructed from M-theory on singular Calabi-Yau threefolds \cite{Morrison:1996xf, Intriligator:1997pq, Jefferson:2017ahm, Xie:2017pfl, DelZotto:2017pti, Jefferson:2018irk, Bhardwaj:2018yhy, Bhardwaj:2018vuu, Closset:2018bjz,  Apruzzi:2019vpe, Apruzzi:2019opn,  Apruzzi:2019enx,  Bhardwaj:2019ngx, Bhardwaj:2019fzv, Saxena:2019wuy, Bhardwaj:2019xeg, Apruzzi:2019kgb}. The advantage of all these constructions is that they provide a systematic approach to study theories with 8 supercharges, within one framework, which often allows extracting highly non-perturbative properties of the theories, such as enhanced flavor symmetries at the conformal fixed points (e.g. see \cite{Heckman:2013pva, Heckman:2015bfa} in 6d and \cite{Apruzzi:2019vpe, Apruzzi:2019opn,  Apruzzi:2019enx, Apruzzi:2019kgb} in 5d). Furthermore, the geometry underlying these string theoretic constructions encodes the spectrum of BPS states and information about the moduli spaces, such as tensor branches, or Higgs and Coulomb branches, in terms of geometric moduli spaces of the singular Calabi-Yau threefolds. 
A complementary approach to the geometric one in 5d are the five-brane webs, which cover a large class of 5d SCFTs and their non-perturbative properties \cite{Aharony:1997ju,Aharony:1997bh,DeWolfe:1999hj, Brandhuber:1999np, Bergman:2012kr, Bergman:2015dpa,
  Zafrir:2015rga, Zafrir:2015ftn, Ohmori:2015tka,Hayashi:2017btw,Hayashi:2018bkd, Hayashi:2015fsa, Hayashi:2018lyv, Hayashi:2019jvx}.

A natural question is whether a similar classification exists for 4d $\mathcal{N}=2$ SCFTs.
In 4d, theories with 8 supercharges are of course very well-studied based on the seminal papers \cite{Seiberg:1994aj, Seiberg:1994rs}. Since the Seiberg-Witten solution, many new (strongly-coupled) theories have been discovered (see e.g. \cite{Argyres:1995jj, Eguchi:1996vu, Minahan:1996fg, Minahan:1996cj, Argyres:2007tq, Cecotti:2010fi, Xie:2015rpa} and earlier work \cite{Sohnius:1981sn, Howe:1983wj}). In particular the so-called class $\mathcal{S}$ construction initiated in \cite{Gaiotto:2009we} has significantly expanded the landscape of $\mathcal{N}=2$ SCFTs in 4d. Although by now we have several tools to study 4d theories with 8 supercharges, we are still missing a complete classification. 
  Recently in a series of papers a classification program was proposed, studying directly the Coulomb branch (CB) geometries of such theories \cite{Argyres:2015ffa, Argyres:2015gha, Argyres:2016xmc, Argyres:2016yzz, Caorsi:2018zsq}. In particular the authors propose a classification of rank one 4d $\mathcal{N}=2$ SCFTs from this approach, and some preliminary results on higher rank extensions. Alternatively, some of these rank one theories have been constructed from twisted torus compactification of 6d/5d SCFTs \cite{Ohmori:2018ona}. As this analysis is entirely developed from a bottom-up point of view, making no {direct} reference to a string theoretic construction, it is natural to ask whether this classification approach can be reproduced from a geometric setup, and ideally extended to higher rank. 

The goal of this paper is to fill this gap and provide a string-theoretic realization of all the rank one theories in \cite{Argyres:2016yzz} (in particular the table 1 in this reference) modulo discrete gauging, and furthermore propose a higher rank generalization. The discrete gaugings can be implemented field theoretically, or in a string theoretic way as in \cite{Harvey:2007ab,Bourget:2019phe, Arias-Tamargo:2019jyh}. The framework we consider is D3-branes probing singularities in type IIB string theory, including quotients that act on the axio-dilaton $\tau$, i.e. S-folds  \cite{Garcia-Etxebarria:2015wns, Aharony:2016kai}. To obtain all rank one SCFTs in 4d with 8 supercharges, we have to slightly extend the notion of S-folds to quotients that preserve $\mathcal{N}=2$ on a probe D3-brane, and also include 7-branes in the background of the S-folds. This approach has several advantages: firstly, it provides a straight-forward generalization to higher rank theories by considering stacks of D3-branes, which are more difficult to study from the Coulomb branch picture. Secondly, we will be able to easily derive the Higgs branch flows from the brane picture, which again, is more challenging from a purely field-theoretic approach. 
From a string theoretic point of view, these S-folds are new backgrounds, which are worth studying in their own right. 
They fall into the class of Type IIB backgrounds where there is a non-trivial action or variation of the coupling, and are thus broadly speaking part of F-theory backgrounds. Probing such backgrounds with D3-branes results in 4d theories with specific values of $\tau$ \cite{Garcia-Etxebarria:2015wns, Aharony:2016kai}, or varying $\tau$ (see \cite{Martucci:2014ema, Assel:2016wcr, Lawrie:2018jut}).

This paper is organized as follows. We begin with a general discussion in section \ref{sec:back} of $\mathcal{N}=2$ S-folds in the presence of 7-branes and the various compatibility constraints. This is followed by an analysis of the quotients of the associated Kodaira singularities in section \ref{sec:quotients}. With this setup in place, we reproduce the classification of rank one 4d $\mathcal{N}=2$ SCFTs in section \ref{sec:RankOne}. In section \ref{sec:ChargesFluxes} we compute the D3-brane charges in the presence of S-fold plus 7-branes, and discuss discrete fluxes, and the necessity to turn these on. 
Finally, in section \ref{sec:CC} we compute the $a$ and $c$ central charges of the 4d theories, using insights from holography, and discuss the extension to higher rank theories.


\section{S-folds with 7-branes} 
\label{sec:back}

We consider Type IIB string theory on an S-fold, which is a generalized orientifold that also acts on the 
axio-dilaton $\tau = C_0 + e^{-\phi} i$ by an S-duality, thereby fixing it to a specific value \cite{Aharony:2016kai, Garcia-Etxebarria:2015wns}. 
The standard application of S-folds is to 4d $\mathcal{N}=3$ theories. We extend this notion to $\mathcal{N}=2$ preserving backgrounds. We apply this 
to construct all rank one 4d $\mathcal{N}=2$ SCFTs and provide a systematic generalization to a class of higher rank theories.
The main idea is to combine $\mathcal{N}=3$ S-folds with 7-branes and to probe these configurations with D3-branes. 

\subsection{4d $\mathcal{N}=3$ and $\mathcal{N}=2$ S-folds}

Consider Type IIB on 4d flat Minkowski space times $\mathbb{C}^3$, which is orbifolded by the spacetime action that multiplies each complex coordinate $z_i$ by a phase as follows
\be\label{rot1}
\rho\equiv \text{diag} (e^{i\Psi_1},e^{i\Psi_2},e^{i\Psi_3})  \,.
\ee
Placing a D3-brane transverse to the $\mathbb{C}^3$ induces an action on the supercharges by the phase rotations in the $SU(4)$ R-symmetry directions, 
which act on the supercharges $Q_i$ of the $\mathcal{N}=4$ SYM theory living on the D3-brane worldvolume by
\begin{widetext}
\be \label{eq:gensu4}
M_\rho= \text{diag}\left(e^{i\frac{(\Psi_1+\Psi_2+\Psi_3)}{2}},e^{i\frac{(\Psi_1-\Psi_2-\Psi_3)}{2}},e^{i\frac{(\Psi_2-\Psi_1-\Psi_3)}{2}},e^{i\frac{(\Psi_3-\Psi_1-\Psi_2)}{2}}\right)  \,.
\ee
\end{widetext}
For a similar analysis see also \cite{Borsten:2018jjm}. 
We would like to determine all such spacetime actions, such that  when supplemented with 7-branes on the D3-brane world-volume $\mathcal{N}=2$ supersymmetry is preserved.

\begin{table*}
\centering
\begin{tabular}{|c|c|c|c|c|}
\hline
Kodaira Type & $G$ & $\Delta_7$ & Weierstrass model & Value of $\tau$ \\
\hline 
$II$ &$H_0$ & $\frac{6}{5}$ & $y^2=x^3+c_{4/5}x+z$ & $e^{\pi i/3}$\\
$II$ &$H_1$ & $\frac{4}{3}$ & $y^2=x^3+xz+c_{2/3}z+M_2$ & $e^{\pi i/2}$\\ 
$IV$ &$H_2$ & $\frac{3}{2}$ & $y^2=x^3+z^2+M_3+x(c_{1/2}z+M_2)$ & $e^{\pi i/3}$\\
$I_0^*$ &$D_4$ & 2 & $y^2=x^3+x(\tau z^2+M_2z+M_4)+ z^3+\widetilde{M}_4z+M_6$ & $\tau$ \\
$IV^*$ &$E_6$ & 3 & $y^2=x^3+z^4+\sum_{i=2}^4M_{3i}z^{4-i}+x\left(\sum_{i=0}^2M_{2+3i}z^{2-i}\right)$ & $ e^{\pi i/3}$\\
$III^*$ &$E_7$ & 4 & $y^2=x^3+x(z^3+M_8z+M_{12})+\sum_{i=0}^4M_{2+4i}z^{4-i}$ & $e^{\pi i/2}$\\ 
$II^*$ &$E_8$ & 6 & $y^2=x^3+z^5+\sum_{i=2}^5M_{6i}z^{5-i}+x\left(\sum_{i=0}^3M_{2+6i}z^{3-i}\right)$ & $e^{\pi i/3}$\\
\hline
\end{tabular}
\caption{The seven Kodaira singularities {with fixed axio-dilaton values} with the corresponding Weierstrass models. We have included explicitly all the deformations including the relevant couplings $c_i$ and mass parameters $M_i$ (where $i$ denotes the scaling dimension) of the corresponding four-dimensional theory living on a probe D3-brane. Only in the $D_4$ case the axio-dilaton (i.e. the 4d gauge coupling) is not frozen to a specific value. $\Delta_7$ denotes the scaling dimension of the CB operator $z$. \label{7branes}}
\end{table*}

The main focus of our paper is the study of the so-called S-folds, where the action on the D3-brane induced by (\ref{rot1}) is accompanied by an $\text{SL}(2,\mathbb{Z})$ transformation. 
The allowed transformations in $\text{SL}(2,\mathbb{Z})$ are of the form $\bbZ_k$ with $k=2,3,4,6$ and it is possible to implement them only if the axio-dilaton $\tau$ is frozen to the following values (there is no restriction on $\tau$ for $k=2$)
\be\label{frozen}\begin{array}{c|c}
\bbZ_k & \text{Value of $\tau$} \\
\hline 
k=2 & \tau \\
k=3 & e^{\pi i/3}\\
k=4 & e^{\pi i/2}\\ 
k=6 & e^{\pi i/3}\\
\end{array}\,.
\ee
Assuming the axio-dilaton is fixed at the required value, the supercharges transform under the S-transformation in $\text{SL}(2, \mathbb{Z})$ as follows \cite{Kapustin:2006pk}
\be \label{eq:SL2Zrot}
S:\qquad Q_i\rightarrow e^{-i\frac{\pi}{k}}Q_i \,.
\ee
If this transformation is accompanied by a rotation (\ref{rot1}) of the form 
\be
\Psi_1=\Psi_2=-\Psi_3=\frac{2\pi}{k} \,,
\ee
three out of the four supercharges are preserved by the quotient and in this way we can engineer $\mathcal{N}=3$ theories in four dimensions. These are $\mathcal{N}=3$-preserving S-folds and have been discussed in detail in \cite{Garcia-Etxebarria:2015wns, Aharony:2016kai}. 

The notion of S-folds can be extended to preserve less supersymmetry, e.g. $\mathcal{N}=2$ S-folds, for which we simply require that two of the phases in $M_\rho$ agree, and are equal opposite to the action of $S$, \eqref{eq:SL2Zrot}, for some value of $k$. Note that of course there are more general spacetime actions than (\ref{rot1}), and the most general setup would leave invariant an $SU(2) \times U(1)$ subgroup of the $SU(4)$, and does not necessarily have to be diagonal. However for our purpose of studying rank one 4d $\mathcal{N}=2$ theories we will require the action to be diagonal. 
The conditions for the $\mathcal{N}=2$ S-folds are (without loss of generality preserving the first two supercharges)
\be \label{eq:N2Sfold}
\Psi_2 = - \Psi_3 \; \text{mod} \; 2 \pi n \,,\qquad 
\Psi_1 = {2 \pi  \over k} \,.
\ee
{Note that we will refer to S-fold backgrounds of this type as $\mathcal{N}=2$ S-folds, but more generally, as we will see in the next section, also to all combinations of S-fold actions with 7-branes that preserve $\mathcal{N}=2$ supersymmetry overall.}

\subsection{S-folds with 7-branes}

{We now turn to generalizing the S-fold backgrounds in Type IIB to include 7-branes, which are placed at the locus $z_1=0$ inside $\mathbb{C}^3$.
 The supercharges preserved by the 7-branes are the ones which have eigenvalue $+1$ with respect to the generator of the rotation in the complex plane $\mathbb{C}_{z_1}$, parametrized by $z_1$. 
We note that this spacetime projection is compatible with the $\mathcal{N}=2$ preserving S-fold, \eqref{eq:N2Sfold}. }

Next we include the $\text{SL}(2,\mathbb{Z})$-action on the supercharges: this is specified by an equal phase rotation, \eqref{eq:SL2Zrot}. We can observe that considering the 7-brane action together with \eqref{eq:SL2Zrot} and \eqref{eq:gensu4}, we get that the first two phases in \eqref{eq:gensu4} are canceled by the $\text{SL}(2,\mathbb{Z})$ action in \eqref{eq:SL2Zrot}, and still preserved by the $z_1$ spacetime rotation due to the 7-branes. In total this results in 8 preserved real supercharges. Additionally, we need to consider the action of the 7-brane on the axio-dilaton. Since the S-transformation \eqref{eq:SL2Zrot} can be performed only when $\tau$ is fixed to one of the values in \eqref{frozen}, we must consider 7-branes which are compatible with this constant value of the axio-dilaton. {In \cite{Gukov:1998kt} a related construction was considered, namely 7-branes with constant $\tau$ wrapping an orbifold singularity, such that the background is a product of two non-compact K3-surfaces. In that case, the theories engineered by probing the geometry with one or more D3-branes have always rank bigger than one. These are other example of $N=2$ background that combine 7-branes with orbifolds, which we will not consider further here.}

Our goal is therefore to construct  $\mathcal{N}=2$-preserving {backgrounds that combine S-folds, which are quotients acting on spacetime and axio-dilaton,} with a stack of 7-branes at a fixed axio-dilaton value. The latter fall into the Kodaira classification of singularities as we will now review.

The list of relevant 7-branes are reported in table \ref{7branes}.
In the first column of table \ref{7branes} the singularity type are shown, whereas in the third column we have indicated the scaling dimension $\Delta_7$ of the corresponding Coulomb branch (CB) operator. Each stack is a bound state of 
\be
n_7={12 \left({\Delta_7-1\over \Delta_7}\right)}
\ee
{mutually} nonlocal 7-branes, whose presence creates a ${2\pi \over \Delta_7}$ deficit angle in the transverse plane. The axio-dilaton is fixed to the values in the last column.
These backgrounds can be described in the context of F-theory by compactifications on $\mathbb{R}^8\times \text{K3}$, where the K3 is elliptic, with non-compact base $\mathbb{C}$, and defined by the Weierstrass model in table \ref{7branes}. We incorporate the $\mathbb{Z}_k$ S-fold by orbifolding a $\mathbb{C}^2$ inside $\mathbb{R}^8$ and accompanying this with a $\mathbb{Z}_k$-quotient of the K3. The latter involves a $\mathbb{Z}_k$ action on the base of the elliptic fibration, which is identified with the Coulomb branch of the four dimensional theory living on the probe D3-brane. The CB operator of the resulting theory will have dimension $k\Delta_7$. This transformation can be seen as a $\mathbb{Z}_{k\Delta_7} \subset U(1)_R$ action, which corresponds to a phase rotation on the supercharges. In order to preserve supersymmetry, we should accompany this with a transformation $\mathbb{Z}_{k\Delta_7}\subset \text{SL}(2,\bbZ)$. We therefore conclude that 
$k\Delta_7=2,3,4$ or $6$, and we find the following six possibilities:
\begin{itemize}
\item  $k=2$ S-fold with 7-branes of type $H_2$, $D_4$ and $E_6$,
\item $k=3$ S-fold with 7-branes of type $H_1$ and $D_4$,
\item $k=4$ S-fold with a 7-brane of type $H_2$.
\end{itemize}
Notice that the $\mathbb{Z}_{k\Delta_7}\subset \text{SL}(2,\mathbb{Z})$ quotient can be performed only if the axio-dilaton is frozen at the required value, as reported in (\ref{frozen}) and this should of course be compatible with the value of the axio-dilaton associated with the given stack of 7-branes (as in table \ref{7branes}). Remarkably, all the six options listed above pass this consistency condition. 

\begin{table*}
$$
\begin{array}{|c|c||c|c|c|c|}
\hline 
$G$ &\Delta_7 & k=1 & k=2  & k=3 &k=4\\ \hline
\hline 
E_8 &6& \csc   [II^*,E_8]  & & &  \\
\hline 
E_7&4 & \csc [III^*,E_7] & & &  \\
\hline
E_6&3 & \csc [IV^*,E_6] & \csc [II^*,C_5]  & &  \\
\hline  
D_4&2 & \csc [I_0^*,D_4]  &\csc [III^*,C_3 C_1]  & \csc [II^*,A_3{\!\rtimes}\Z_2] &  \\
\hline  
H_2&3/2 &\csc [IV,H_2]  &  \csc [IV^*,C_2 U_1] & &\csc [II^*,A_2{\!\rtimes}\Z_2]   \\
\hline
H_1&4/3& \csc [III,H_1] & &  \csc [III^*,A_1U_1{\!\rtimes}\Z_2]  &  \\
\hline    
H_0&6/5 & \csc [II,H_0]& &  & \\
\hline    
\varnothing&1&& \csc \blue{[I_0^*,C_1 \chi_0]} &\csc \green{[IV^*,U_1]}  & \csc \green{[III^*,U_1{\!\rtimes}\Z_2]}  \\
\hline    
&& & & & \csc [IV^*_1, \varnothing]  \\
\hline   
\end{array}$$
\caption{S-fold with 7-branes, where $k$ is the S-fold action $\mathbb{Z}_k$ and 
{$G$ the 7-brane type. In the following we will denote these models by $ \langle G, \mathbb{Z}_k\rangle$.}
$\Delta_7$ indicates the 7-brane type realized by a Weierstrass model as in table \ref{7branes}. {Each entry in the table specifies the Kodaira type and the global symmetry, which makes contact with the notation in \cite{Argyres:2016yzz}.}
Models shown in black are $\mathcal{N}=2$,  blue is an $\mathcal{N}=4$ theory and the green entries are $\mathcal{N}=3$ theories. \label{tab:AllMods}}
\end{table*}

Before discussing in detail the quotients of Kodaira singularities, let us pause to explain why we think of these backgrounds, {which we will also refer to as $\mathcal{N}=2$ S-folds,} as $\mathcal{N}=3$ S-folds and 7-branes combined: if we probe the background with a D3-brane and we move it along the $\mathbb{C}^2$ corresponding to the last two entries in (\ref{rot1}), away from the fixed point of the $\bbZ_k$ action, the probe D3-brane  does not perceive any longer the presence of the quotient and therefore sees only the Kodaira singularity, i.e. the 7-brane. In this sense the background includes a 7-brane wrapping the same $\mathbb{C}^2$. If on the other hand we deform the resulting singularity by activating the deformation parameters of the original 7-brane which survive the $\bbZ_k$ quotient, we are left with a $\bbZ_k$ $\mathcal{N}=3$ S-fold only, as we will show in detail in the next section. We therefore recover both building blocks.

We can organize the consistent S-folds in the presence of 7-branes, labelled by their singularity type ($E_8, E_7, E_6, D_4, H_2, H_1, H_0, \varnothing$) according to the nontrivial $\bbZ_{k}$ action of the quotient on the CB. We will give evidence for the presence of  the possibilities in table \ref{tab:AllMods}. {These models will be denoted by 
\be
\langle G, \mathbb{Z}_k \rangle \,,
\ee
where $G$ labels the 7-brane singularity type and $\mathbb{Z}_k$ the S-fold. 
}
In the next section we will show that all models in a column with fixed $k$ are related by mass deformations, i.e. RG-flows, by considering the quotients of the corresponding Kodaira fibers.

\section{Quotients of Kodaira  Singular Fibers}
\label{sec:quotients}

We now describe in detail the quotient of the Kodaira singularities. A similar quotient in F-theory was considered also in \cite{Tachikawa:2015wka} to describe discrete three-form flux in M-theory.
A complementary point of view is to consider the Seiberg-Witten curve for the theories in table \ref{tab:AllMods} as performed in \cite{Argyres:2016yzz}. Here we will work on the level of the Kodaira singular fiber and constrain the {quotients} by requiring the invariance of the holomorphic 2-form. 

Let us start by explaining the general idea: If we write the Kodaira singularities in table \ref{7branes} in the form $W(x,y,z)=0$, the corresponding holomorphic two-form {of the associated K3-surface is}
\be\label{2holo}\Omega_2=\frac{dz\wedge dx\wedge dy}{dW} \,.
\ee 
We want the $\bbZ_k$ quotient to act on $z$ as $z\rightarrow e^{2\pi i/k}z$ and therefore we can introduce the invariant coordinate $U=z^k$. We then assign a transformation law to $x$ and $y$ in such a way that $y^2$ and $x^3$ transform in the same way and $\Omega_2$ is invariant under the quotient. We also introduce the corresponding invariant coordinates $X$ and $Y$, which are obtained by rescaling $x$ and $y$ by suitable powers of $z$, and require that $\Omega_2$ can be written in terms of $X$, $Y$ and $U$ only. These requirements imply that the invariant coordinates are 
\be\label{coordch} X=xz^{2k-2}\,,\qquad Y=yz^{3k-3}\,.
\ee 
Furthermore, the holomorphic two-form in the new coordinates reads 
\be\label{holonew}\Omega_2=\frac{dU\wedge dX\wedge dY}{dW(X,Y,U)}\,.
\ee 
In the previous formula we have implicitly assumed that the Kodaira singularity can be rewritten in terms of the invariant coordinates only. As we will see momentarily, this is possible only for certain values of $k$ (reproducing the list of allowed S-fold/7-branes found earlier, see table \ref{tab:AllMods}) and only when suitable constraints on the deformations of the singularity are imposed. All the quotients which are not discussed explicitly below are not consistent. 
{For each of the consistent models we determine the scaling dimension $D$ of Coulomb branch operators for the theories on the D3-branes probing the singularity.}

\subsection{$\mathbb{Z}_2$-Quotient of Type $IV^*$ Fibers}

We start with the $IV^*$ singularity (or 7-brane of type $E_6$) including mass deformations
\be\ba
IV^* : 
\quad  y^2 &= x^3 + x (z^2  M_2 + z M_5 + M_8) \cr 
&+ z^4 + M_6 z^2 + M_9z + M_{12} \,.
\ea\ee
In order to perform the  $\mathbb{Z}_k$ quotient we consider the change of variables (\ref{coordch})
\be\label{ZaQuot}
Y= y z^{3 (k-1)} \,,\qquad 
X= x z^{2 (k-1)} \,,
\ee
which brings the curve to the form
\be\ba
Y^2 = &X^3 + X (z^{4k-2}  M_2 + z^{4k-3} M_5 + z^{4k-4}M_8) \cr 
&+ z^{6k-2} +  z^{6k-4}M_6 +z^{6k-5} M_9 + M_{12}z^{6k-6}\,.
\ea\ee
Even if we turn off all the mass deformations we still have the leading term of the singularity $z^{6k-2}$ and we require this to be an integer power of the invariant coordinate $U=z^k$. Clearly this is the case only for $k=2$ and therefore we restrict to this case from now on. Notice that we should set $M_5=M_9=0$ since the corresponding terms cannot be written in terms of $U$. We are therefore left with the $II^*$ singularity 
\be\ba
\label{newe6}
II^*:\qquad Y^2 = &X^3 +X \left( U^3M _2+U^2M_8\right) \cr 
&+ U^{5}+U^4M_6+U^3M_{12} \,.
\ea\ee
We can now determine the scaling dimension of the new Coulomb branch operator $U$ by imposing homogeneity of the curve and the condition that the dimension of the SW differential (and therefore the dimension of $\Omega_2$ as well) is equal to one. From (\ref{holonew}) we find 
\be\ba
&D(Y)=15, \;\; D(X)=10,\;\; D(U)=6,\;\; \cr 
&D(M_2)=2,\;\; D(M_6)=6,\;\; D(M_8)=8,\;\; D(M_{12})=12 \,.
\ea\ee
The geometry (\ref{newe6}) can be effectively used to study deformations of the theory (or equivalently deformations of the stack of 7-branes in the presence of the $\bbZ_2$ S-fold). If for instance we turn on $M_2$ we land on the following type $III^*$ singularity
\be\label{defe6}
III^*:\qquad Y^2 = X^3 +X \left( U^3+U^2M_8\right)+ U^4M_6+U^3M_{12}\,.
\ee
We will recover this geometry later.

\subsection{$\mathbb{Z}_2$- and $\mathbb{Z}_3$-Quotients of Type $I_0^*$ Fibers}

The Weierstrass model for the $I_0^*$ singularity (or 7-brane of type $D_4$) is 
\be I_0^*:\quad y^2=x^3+x(\tau z^2+M_2z+M_4)+ z^3+\widetilde{M}_2z^2+\widetilde{M}_4z+M_6\,.
\ee 
Of course by shifting $z$ we can get rid of $M_2$ or $\widetilde{M}_2$: only one of them is a physical parameter and in table \ref{7branes} we have removed $\widetilde{M}_2$. Here we prefer keeping both parameters for later convenience. 

If we now consider as before the change of variables (\ref{coordch}), we find 
\be\ba
\label{ccd4} 
Y^2=&X^3+X(\tau z^{4k-2}+z^{4k-3}M_2+z^{4k-4}M_4)\cr 
&+ z^{6k-3}+z^{6k-4}\widetilde{M}_2+z^{6k-5}\widetilde{M}_4+z^{6k-6}M_6 \,.
\ea\ee
If we now impose that the leading term $z^{6k-3}$ is an integer power of $U$, we conclude that $k=3$ and we should set to zero $\tau$, $M_4$, $\widetilde{M}_2$ and $\widetilde{M}_4$ since the corresponding terms cannot be rewritten in terms of invariant coordinates. We therefore obtain the $II^*$ singularity 
\be\label{d4z3} II^*:\qquad Y^2=X^3+XU^{3}M_2+ U^{5}+U^{4}M_6\,.
\ee 
Again we can determine the scaling dimension of $U$ by imposing that $\Omega_2$ has scaling dimension 1. We find 
\be\ba
 D(Y)&=15\,, \;\; D(X)=10\,, \;\; D(U)=6\,, \cr 
 D(M_2)&=2\,, \;\; D(M_6)=6\,.
\ea\ee 
Notice that the value of the axio-dilaton is automatically frozen at the value $\tau=e^{\pi i/3}$ as required for a singularity of type $II^*$. As in the previous case we can analyze mass deformations: if we turn on the mass parameter $M_2$ we find again a $III^*$ singularity of the form 
\be\label{defd41} III^*:\qquad Y^2 = X^3 +X U^3+ U^4M_6\,.\ee 
We will exploit this result later. 

This case is special as it is the only 7-brane for which the axio-dilaton is not frozen to a specific value. Correspondingly, there are two leading singular terms and there is another case we should consider: we can require that the term $\tau z^{4k-2}$ is expressible in terms of $U$, which implies $k=2$. This time we should discard the term $z^{6k-3}$ since it cannot be expressed in terms of $U$ and the axio-dilaton is frozen to the value $\tau=i$. We should also set $M_2=\widetilde{M}_4=0$. We end up with the $III^*$ geometry 
\be\label{newd4} III^*:\qquad Y^2=X^3+X(U^3+U^2M_4)+U^4\widetilde{M}_2+U^3M_6\,.\ee 
We have removed the parameter $\tau$ since it can now be reabsorbed with a rescaling of the coordinates and a redefinition of the parameters. Notice that (\ref{newd4}) is precisely the geometry (\ref{defe6}). We therefore conclude that the geometry corresponding to a 7-brane of type $E_6$ combined with a $\bbZ_2$ S-fold can be deformed to the geometry corresponding to a 7-brane of type $D_4$ combined with a $\bbZ_2$ S-fold. Analogously to the previous cases we can analyze mass deformations: when we turn on $\widetilde{M}_2$ we find the $IV^*$ singularity 
\be\label{defd42} IV^*:\qquad Y^2 = X^3 +X U^2M_4+ U^4+U^3M_6\,.
\ee

\subsection{$\mathbb{Z}_2$- and $\mathbb{Z}_4$-Quotients of Type $IV$ Fibers}

The Type $IV$ singularity (7-brane of type $H_2$) is 
\be
IV:\qquad y^2=x^3+x(zc_{1/2}+M_2)+z^2+M_3\,.
\ee
Performing again the change of variables (\ref{coordch}) we find 
\be\label{h2quot} Y^2=X^3+X\left(z^{4k-3}c_{1/2}+z^{4k-4}M_2\right)+z^{6k-4}+z^{6k-6}M_3\,,
\ee 
and demanding again that the leading term $z^{6k-4}$ can be written in terms of $U=z^k$ we easily see that there are two options for the quotient:  $\mathbb{Z}_2$ and $\mathbb{Z}_4$.

Consider first the $\mathbb{Z}_2$ quotient. From (\ref{h2quot}) we conclude that we have to set $c_{1/2}=0$ and the resulting singularity is of type $IV^*$:
 \be
 IV^*:\qquad  Y^2=X^3+XU^2M_2+U^4+U^3M_3\,.
 \ee
Demanding that the holomorphic two-form (\ref{2holo}) has dimension one we find $D(U)=3$. Notice that this is precisely the geometry we find by deforming the $\bbZ_2$ quotient of the $D_4$ singularity (\ref{defd42}). Let us now study the deformation: if we turn on $M_3$ (the other option has exactly the same effect) we get a $I_0^*$ singularity which is not further deformable: 
 \be
 I_0^*:\qquad  Y^2=X^3+X\tau'U^2+U^3\,.
 \ee
In this case the parameter $\tau$ (inherited from $M_2$) can be easily checked to be dimensionless and plays the role of a marginal coupling in the 4d theory living on the D3-brane probe: we can easily see that the j-invariant is not constrained in the above geometry. What we find here is just the Weierstrass model for the $\bbZ_2$ S-fold without any 7-branes!

The analysis for the $\mathbb{Z}_4$ quotient is similar: in this case we should turn off both $c_{1/2}$ and $M_3$, resulting in the $II^*$ geometry
\be
II^* :\qquad Y^2 = X^3  +  X U^3 M_2 + U^5\,, 
\ee
with the scaling dimensions 
\be
D(Y)=15\,\; D(X)=10\,, \; D(U)=6\,,\; D(M_2)=2\,.
\ee
 By turning on the deformation parameter $M_2$ we land on the frozen $III^*$ singularity
\be III^* :\qquad Y^2 = X^3  +  X U^3, \ee
which we interpret as the Weierstrass model associated with a $\bbZ_4$ S-fold.

\subsection{$\mathbb{Z}_3$-Quotient of Type $III$ Fibers}

Consider the Type $III$ singularity (i.e. 7-brane of type $H_1$), with mass deformation $M_2$
\be\label{IIISing}
III: \qquad y^2 = x^3 + xz  + zc_{2/3}+M_2\,.
\ee
After the change of coordinates (\ref{coordch}) we get 
\be\label{h1quot} Y^2=X^3+Xz^{4k-3}+z^{6k-5}c_{2/3}+z^{6k-6}M_2\,,\ee 
and by demanding that $z^{4k-3}$ can be written as an integer power of $U$ we conclude that $k=3$. We should then set to zero $c_{2/3}$ and we obtain a $III^*$ singularity: 
\be\label{h1z3} III^*: \qquad  Y^2=X^3+XU^{3}+U^4M_2\,.
\ee 
The dimension of the CB operator $U$ of the 4d field theory living on a probe D3-brane is 4. Notice that this is precisely the singularity we found by deforming the 7-brane of type $D_4$ combined with a $\bbZ_3$ S-fold (\ref{defd41}). The mass deformation of (\ref{h1z3}) leads to a frozen $IV^*$ singularity which we associate with a $\bbZ_3$ S-fold: 
$$ Y^2=X^3+U^4.$$


\section{Rank one 4d $\mathcal{N}=2$ SCFTs from S-folds}
\label{sec:RankOne}

In this section we will discuss the physics of the four-dimensional $\mathcal{N}=2$ theory living on a D3-brane probing the backgrounds we have studied in section \ref{sec:back} and provide evidence for the identification of these theories with the models recently discussed by \cite{Argyres:2016yzz}. We will identify them with the  models in \cite{Argyres:2016yzz} that do not correspond to discrete gaugings.

\subsection{$\mathcal{N}=2$ S-folds and Rank One Theories}

In section \ref{sec:quotients} we have seen that our backgrounds can be deformed to $\mathcal{N}=3$ preserving S-folds according to the following pattern, within the table \ref{tab:AllMods}
\begin{itemize}
\item $k=2$: \\
$\langle\bbZ_2,E_6\rangle\longrightarrow \langle\bbZ_2,D_4\rangle\longrightarrow \langle\bbZ_2,H_2\rangle\longrightarrow \bbZ_2$ S-fold. 
\item $k=3$: \\ 
$\langle\bbZ_3,D_4\rangle\longrightarrow \langle\bbZ_3,H_1\rangle\longrightarrow \bbZ_3$ S-fold.
\item $k=4$: \\
 $\langle\bbZ_4,H_2\rangle\longrightarrow \bbZ_4$ S-fold. 
\end{itemize}
Of course we know that the endpoints of these flows have enhanced $\mathcal{N}=3$ (actually $\mathcal{N}=4$ in the $\bbZ_2$ case) supersymmetry.
From the perspective of the field theory living on a probe D3-brane, the deformations we have discussed in the previous section correspond to mass deformations. The question now is: which $\mathcal{N}=2$ theories flow to rank one models with enhanced supersymmetry upon a mass deformation according to the pattern we have just described? The answer is given in terms of the theories appearing in table \ref{tab:AllMods}. We are therefore led to the following identification between theories living on a D3-brane probing the $\mathcal{N}=2$ S-folds and known rank-one theories 
\be\label{ident4d}
\begin{array}{c|l} 
\text{S-fold with 7-branes: } \langle G, \mathbb{Z}_k \rangle& \ \text{[Kodaira Type, $G_F$]}\cr\hline 
{\langle E_6, \bbZ_2\rangle}  \qquad &\qquad [II^* ,C_5] \cr 
{\langle D_4, \bbZ_2 \rangle }  \qquad &\qquad [III^*,C_3 C_1] \cr 
\langle H_2, \bbZ_2 \rangle  \qquad &\qquad [IV^*,C_2U_1] \cr 
\langle D_4, \bbZ_3 \rangle  \qquad &\qquad [II^*,A_3{\!\rtimes}\mathbb{Z}_2] \cr 
 \langle H_1, \bbZ_3 \rangle  \qquad &\qquad [III^*,A_1U_1{\!\rtimes}\mathbb{Z}_2] \cr 
 \langle H_2, \bbZ_4 \rangle  \qquad &\qquad  [II^*,A_2{\!\rtimes} \mathbb{Z}_2] \,.
\end{array}
\ee
At this stage it is worth making a remark about the global symmetry: the Weierstrass models for the quotient singularities we have derived in section \ref{sec:quotients} do not correspond to the SW-curve  of the corresponding 4d theories with all the mass parameters turned on. If we denote by $r$ the rank of the actual global symmetry of the field theory in the Weierstrass model we see only $r-1$ mass parameters (which are in general polynomials in the Casimirs of the full global symmetry). The explanation for this apparent mismatch is rather clear: the $\mathcal{N}=3$ theories we land on once we have activated all possible deformations are not the endpoints of the RG-flow, since for every  $\mathcal{N}=3$ theory we can turn on an $\mathcal{N}=2$-preserving mass deformation. The underlying S-fold background on the other hand is a terminal singularity and cannot be deformed, so we conclude that the corresponding mass parameter cannot be interpreted as a deformation of the geometry. Taking this caveat into account, we see that the rank of the global symmetry predicted by our geometric construction perfectly agrees with the known result. In section \ref{hbflow} we will discuss another remarkable piece of evidence for our claim by considering the Higgs branch flows. 

\subsection{The Enhanced Coulomb Branch \label{sec:ECB}} 

As we have seen our proposal for the 4d theories living on a D3-brane probe is strongly supported by the pattern of deformations of our backgrounds. On the other hand, it automatically leads to a sharp prediction for the effective four dimensional theory we get by compactifying F-theory on an $\mathcal{N}=2$-preserving S-fold (without adding any D3-branes). This can be seen as follows: suppose that we move the probe D3-brane away from the 7-brane. Field theoretically this corresponds to moving away from the origin of the Coulomb Branch. Since the probe is now at a smooth point of spacetime (away both from the fixed point of our quotient and from the 7-brane), the low-energy effective theory includes the modes on the D3-brane, namely a free vectormultiplet and a free hypermultiplet and a second sector consisting of the 4d theory one gets by compactifying F-theory on an $\mathcal{N}=2$-preserving S-fold. These two sectors are expected to be decoupled from one another in the low energy limit.

On the other hand, we know exactly what is the low-energy effective theory for the rank one models appearing in table \ref{tab1}   at a generic point on the Coulomb Branch: there is a free vectormultiplet and a collection of $h$ free hypermultiplets ($h$ is called the dimension of the enhanced Coulomb Branch). Therefore, if our claim is correct, we conclude just by looking at  table \ref{tab1} that the 4d theory corresponding to an S-fold compactification of F-theory is given by a collection of $k(\Delta_7-1)$ free hypermultiplets (where indeed $k$ is the order of the orbifold and $2\pi/\Delta_7$ is the deficit angle induced by the 7-brane). We do not know how to check this result and it would definitely be interesting to fill in this gap. However, we will be able to provide a nontrivial consistency check for this prediction when we compute the central charges $a$ and $c$ of the field theory on the probe D3-brane: we will see that the central charges match those of the known rank one theories only if the central charges of F-theory compactified on an $\mathcal{N}=2$-preserving S-fold are those of  $k(\Delta_7-1)$ free hypermultiplets.  

\subsection{Higgs Branch Flows}\label{hbflow}

Our proposed identification of the field theories living on a D3-brane probing an $\mathcal{N}=2$-preserving S-fold with the rank one theories in \cite{Argyres:2016yzz}
passes a highly nontrivial consistency check. The motion of the D3-brane within the 7-brane (i.e. along $z_2$ and $z_3$), but away from the S-fold singularity, geometrizes an RG-flow initiated by turning on an expectation value for a Higgs branch operator\footnote{This is simply because this motion does not break the $U(1)_R$ symmetry of the theory which acts on the plane transverse to the 7-brane only.}. The low-energy effective theory in the infrared can be easily predicted by recalling one of the remarks we have made in the previous section: once the D3-brane is away from the S-fold singularity, the probe does not perceive the presence of the S-fold quotient anymore, but it still sees the 7-brane. We therefore conclude that the low-energy effective theory is that of a D3-brane probing the same type of 7-brane in flat space plus the effective four dimensional theory coming from the background, i.e. a collection of $k(\Delta_7-1)$ hypermultiplets in our case. We therefore predict the existence of the following Higgs branch flows
\be
\ba
{[II^* ,C_5]}   &\quad  \longrightarrow\quad  [IV^*,E_6] + 5\;  \text{hypers} \\
 [III^*,C_3 C_1]& \quad\longrightarrow \quad [I_0^*,D_4] +3\;  \text{hypers}  \\ 
 [IV^*,C_2U_1] &\quad\longrightarrow  \quad [IV,A_2]+2\;  \text{hypers}  \\
[II^*,A_3{\!\rtimes}Z_2] &\quad\longrightarrow\quad [I_0^*,D_4]+4\;  \text{hypers} \\ 
[III^*,A_1U_1{\!\rtimes}Z_2]&\quad \longrightarrow\quad [III;A_1]+2\;  \text{hypers}  \\
 [II^*,A_2{\!\rtimes}Z_2] &\quad\longrightarrow\quad  [IV,A_2]+3\;  \text{hypers}  \,.\\
\ea
\ee
It is easy to verify that all these RG flows are consistent with 't Hooft anomaly matching for the unbroken $U(1)_R$ symmetry. 

Remarkably, these RG-flows match those, that are found purely field theoretically in \cite{TBABMMPR}. 
Our geometric setup allows us {to derive and predict}  this nontrivial dynamical fact in a simple and natural way. 

\subsection{Global Symmetries and Mass Parameters}

 The global symmetry of the 4d theory on the D3-brane arises from the gauge symmetry supported on 7-branes and also from isometries of the F-theory background. In the well known case of D3-branes probing flat 7-branes (the case $k=1$ in our notation) for example, the isometries of the background provide a $SU(2)$ (non R) global symmetry which acts on the free hypermultiplet parametrizing the position of the center of mass of the D3-branes along the worldvolume of the 7-branes (and also on the strongly-coupled SCFT if we have multiple D3-branes).

\begin{table*}
\centering
\begin{tabular}{|c|c|c|c|}\hline
Model & $h$ & $a$ &$c$ \cr \hline\hline 
&&&\cr
$[II^*,E_8]$  & $0$ & $\frac{95}{24}$ &$\frac{31}{6}$ \cr 
&&&\cr\hline 
&&&\cr
$[III^*,E_7]$  & $0$ & $\frac{59}{24}$ &$\frac{19}{6}$ \cr 
&&&\cr\hline 
&&&\cr
$[IV^*,E_6]$  & $0$ & $\frac{41}{24}$ &$\frac{13}{6}$ \cr
&&&\cr \hline 
&&&\cr
$[IV,H_2]$  & $0$ & $\frac{7}{12}$ &$\frac{2}{3}$ \cr &&&\cr\hline 
&&&\cr
$[III,H_1]$  & $0$ & $\frac{11}{24}$ &$\frac{1}{2}$ \cr&&&\cr \hline 
&&&\cr
$[III,H_0]$  & $0$ & $\frac{43}{120}$ &$\frac{11}{30}$ \cr&&&\cr \hline 
\end{tabular}
\begin{tabular}{|c|c|c|c|}\hline 
Model & $h$ & $a$ &$c$ \cr \hline \hline
&&&\cr
$[II^*,C_5]$  & $5$ & $\frac{41}{12}$ &$\frac{49}{12}$ \cr&&&\cr \hline 
&&&\cr
$[III^*,C_3C_1]$  & $3$ & $\frac{25}{12}$ &$\frac{29}{12}$ \cr &&&\cr\hline 
&&&\cr
$[IV^*,C_2U_1]$  & $2$ & $\frac{17}{12}$ &$\frac{19}{12}$ \cr &&&\cr\hline 
&&&\cr
$\blue{[I^*_0,C_1 \chi_0]}$  & $1$ & $\frac{3}{4}$ &$\frac{3}{4}$ \cr &&&\cr\hline 
\end{tabular}
\begin{tabular}{|c|c|c|c|}\hline 
Model & $h$ & $a$ &$c$ \cr \hline \hline
&&&\cr
$[II^*,A_3 \rtimes \mathbb{Z}_2]$  & $4$ & $\frac{25}{8}$ &$\frac{7}{2}$ \cr&&&\cr \hline 
&&&\cr
$[III^*,A_2U_1\rtimes \mathbb{Z}_2]$  & $2$ & $\frac{15}{8}$ &$2$ \cr&&&\cr \hline 
&&&\cr
$\green{[IV^*,U_1]}$  & $1$ & $\frac{5}{4}$ &$\frac{5}{4}$ \cr &&&\cr\hline
\end{tabular}
\begin{tabular}{|c|c|c|c|}\hline 
Model & $h$ & $a$ &$c$ \cr \hline \hline
&&&\cr
$[II^*,A_2 \rtimes \mathbb{Z}_2]$  & $3$ & $\frac{71}{24}$ &$\frac{19}{6}$ \cr &&&\cr\hline 
&&&\cr
$\green{[III^*,U_1\rtimes \mathbb{Z}_2]}$  & $1$ & $\frac{7}{4}$ & $\frac{7}{4}$ \cr&&&\cr \hline 
&&&\cr
$[IV^*,\varnothing ]$  & $0$ & $\frac{55}{48}$ & $\frac{25}{24}$ \cr &&&\cr\hline 
\end{tabular}
\caption{Coulomb branch dimensions for 4d $\mathcal{N}=2$ rank 1 theories in the notation of table \ref{tab:AllMods}. Again the color coding is: black are $\mathcal{N}=2$ theories, blue are $\mathcal{N}=4$ and green are $\mathcal{N}=3$ supersymmetric. \label{tab1}}
\end{table*}

As far as the global symmetry coming from 7-branes is concerned, we can easily read off the rank from the Weierstrass quotients we have discussed in 
section \ref{sec:quotients}. The corresponding mass parameters are in one-to-one correspondence with the deformations of the Weierstrass moel, which survive the $\mathbb{Z}_k$ quotient. We therefore have 4 mass parameters for the $\mathbb{Z}_2$ quotient of the $IV^*$ singularity, 3 for the $\mathbb{Z}_2$ quotient of the $I_0^*$ singularity and so on.

In the case of $\mathcal{N}=2$ S-folds we are considering, the isometries of the background include the $U(1)_R$ symmetry of the SCFT, which acts on the plane parametrized by $z_1$, there is also a $SU(2)$ isometry of the quotient $\mathbb{C}^2/\mathbb{Z}_k$, which is identified with the $SU(2)$ R-symmetry, and then a further $U(1)$ (actually enhanced to $SU(2)$ for $k=2$) which is a global (non R) symmetry of the theory. 
This symmetry is of course there also in cases with enhanced supersymmetry (when there are no 7-branes) and we can turn on a corresponding $\mathcal{N}=2$-preserving mass deformation. We should point out that this mass parameter cannot be described by deforming the singularity (which indeed is not deformable in the $\mathcal{N}=3$ case), but rather arises by turning on a $\Omega$-background along directions $z_2$ and $z_3$. This is how $\mathcal{N}=2^*$ theories are engineered in Type IIB (see \cite{Billo:2012st}). If we denote by $\epsilon_2$ and $\epsilon_3$ the $\Omega$-background parameters associated with rotations in the $z_2$ and $z_3$ planes respectively, by setting $\epsilon_2+\epsilon_3=0$ we preserve $\mathcal{N}=2$ supersymmetry and the parameter $\epsilon_3$ is identified with the  $\mathcal{N}=2$-preserving mass\footnote{We thank J. F. Morales for explaining this point to us.}. In the perturbative cases discussed in \cite{Billo:2012st} one can check that stringy modes corresponding to fields charged under the global symmetry acquire a mass proportional to $\epsilon_3$. In our case we claim that, once all the mass parameters appearing in the Weierstrass have been activated and we flow to a model with enhanced supersymmetry,  the theory living on the D3-brane probe we get by further turning on the $\Omega$-background deformation is identified with the mass-deformed version of the underlying $\mathcal{N}=3$ theory (or $\mathcal{N}=4$ in the case $k=2$). 

In summary, the rank of the global symmetry of the theory is obtained just by counting mass deformations in the Weierstrass and adding one (the contribution from $\epsilon_3$). We therefore find {
\be\begin{array}{|c|c|}
\hline
\text{Theory $\langle G, \mathbb{Z}_k\rangle$} & \text{Rank}(G_F) \\
\hline \hline
\langle\bbZ_2,E_6\rangle & 5 \\ 
\hline 
 \langle\bbZ_2,D_4\rangle & 4 \\
\hline 
 \langle\bbZ_2,H_2\rangle & 3 \\
\hline 
 \langle\bbZ_3,D_4\rangle & 3 \\ 
\hline 
 \langle\bbZ_3,H_1\rangle & 2 \\
\hline 
 \langle\bbZ_4,H_2\rangle & 2 \\ 
\hline
\end{array}\ee
}
This is in perfect agreement with the actual rank of the global symmetry of the $\mathcal{N}=2$ SCFT's we associate with our $\mathcal{N}=2$ S-folds, therefore landing further support to our claim.

\section{D3-brane Charges, Discrete Fluxes and Central Charges} 
\label{sec:ChargesFluxes}

In this section we will compute the D3-brane charge of our systems of S-folds plus 7-branes, which will be needed later on when we compute the $a$ and $c$ central charges. As a first step we perform in section \ref{noflux} the computation neglecting the contribution from the torsional flux following the approach of \cite{Bergman:2009zh}. We then incorporate the discrete flux in section \ref{flux}. The outcome of our analysis is that the D3-brane charge of our backgrounds does not depend on the type of 7-brane we introduce and equals that of the underlying S-fold.

\subsection{Computation of the D3-brane Charge}\label{noflux} 

{The induced D3-brane charge is sourced by background D3-branes, fluxes and the contribution to the Euler characteristic from the bulk, meaning the fixed points of the orbifold (the total space being non-compact we do not impose any cancellation of this induced charge) \cite{Bergman:2009zh}. In the absence of fluxes and background D3-branes this is
\be
\epsilon_{D3}=- {\chi_{\text{bulk}}  \over 24 } \,.
\ee
We present two computations of the Euler characteristic.  In appendix \ref{app:D3alt} it is derived in a purely geometric way. In the following we will compute it using the Euler characteristic of orbifolds of $T^8$ obtained in \cite{Font:2004et}.}
 In the following we will analyze in detail three examples: the $\mathbb{Z}_2$ S-fold combined with 7-branes of type $D_4$ and $E_6$ respectively and the $\mathbb{Z}_3$ S-fold combined with 7-branes of type $D_4$. Before entering the details, let us first remind the reader about the following well-known facts regarding orbifolds of $T^2$: 
\begin{itemize}
\item $\faktor{T^2}{\mathbb{Z}_2}$ has four fixed points. 
\item $\faktor{T^2}{\mathbb{Z}_3}$ has three fixed points. 
\item $\faktor{T^2}{\mathbb{Z}_4}$ has two fixed points and two $\mathbb{Z}_2$-{fixed point}s which are interchanged by the residual $\mathbb{Z}_2$. 
\item $\faktor{T^2}{\mathbb{Z}_6}$ has one fixed point, two $\mathbb{Z}_3$-{fixed point}s which are interchanged by the residual $\mathbb{Z}_2$ and three $\mathbb{Z}_2$-{fixed point}s which are interchanged by the residual $\mathbb{Z}_3$.
\end{itemize}
We will use these facts repeatedly below.

\begin{center}
{\it {$\mathbb{Z}_2$ S-fold in the presence of a $D_4$ 7-brane} }
\end{center}

In this case the relevant fourfold can be described by the following $\mathbb{Z}_4$ quotient of $T^2\times\mathbb{C}^3$: 
\begin{equation}\label{d4z2}
\begin{array}{cccc} 
T^2 & \mathbb{C} & \mathbb{C} & \mathbb{C} \\
\hline 
\gamma_4 & \gamma_4^{-1} & \gamma_4^{2} & \gamma_4^{-2} \\
\end{array}
\end{equation} 
where indeed $\gamma_4^4=1$. In order to compute the induced D3-brane charge, we start by considering the same kind of  $\mathbb{Z}_4$ quotient for $T^8$: 
\begin{equation}\label{t8orb}\begin{array}{cccc} 
T^2 & T^2 & T^2 & T^2 \\
\hline 
\gamma_4 & \gamma_4^{-1} & \gamma_4^{2} & \gamma_4^{-2} \\
\end{array}\end{equation}
The Euler characteristic of this fourfold, which we denote as $\faktor{T^8}{\widetilde{\mathbb{Z}}_4}$, was found in \cite{Font:2004et} to be 
$$\chi\left(\faktor{T^8}{\widetilde{\mathbb{Z}}_4}\right)=192.$$ 
The contribution to the Euler characteristic comes from the fixed points under the orbifold action. In our case we have 64 points invariant under the $\mathbb{Z}_4$ action: there are $2^2$ {fixed point}s coming from the first two $T^2$ factors in (\ref{t8orb}), times the 16 points of the remaining $T^4$ which are invariant under the action of $\gamma_4^2$. Around each of these singular points the singularity is equivalent to an orbifold of $\mathbb{C}^4$ of the form 
$$\begin{array}{cccc} 
\mathbb{C} & \mathbb{C} & \mathbb{C} & \mathbb{C} \\
\hline 
\gamma_4 & \gamma_4^{-1} & \gamma_4^{2} & \gamma_4^{-2} \\
\end{array}$$ 
In the following we will denote this space as $\faktor{\mathbb{C}^4}{\widetilde{\mathbb{Z}}_4}$.
The $\mathbb{Z}_2$-{fixed point}s of the first two $T^2$ factors in (\ref{t8orb}) do not contribute: These are invariant under a $\mathbb{Z}_2$ subgroup acting as follows: 
$$\begin{array}{cccc} 
T^2 & T^2 & T^2 & T^2 \\
\hline 
\gamma_4^2 & \gamma_4^{-2} & 1 & 1 \\
\end{array}$$
Around each such point we therefore have the singularity $\mathbb{C}^2/\mathbb{Z}_2\times T^4$, which has vanishing Euler character. Overall, we find 
$$\chi\left(\faktor{T^8}{\widetilde{\mathbb{Z}}_4}\right)=64\chi\left(\faktor{\mathbb{C}^4}{\widetilde{\mathbb{Z}}_4}\right)=192 \longrightarrow \chi\left(\faktor{\mathbb{C}^4}{\widetilde{\mathbb{Z}}_4}\right)=3.$$ 
Since we are ultimately interested in the space (\ref{t8orb}), we should take into account the fact that there are two $\mathbb{Z}_4$-{fixed point}s in $T^2$. Again, the $\mathbb{Z}_2$-{fixed point}s can be ignored since the subgroup which leaves them invariant acts trivially on the last two $\mathbb{C}$ factors in (\ref{t8orb}) and therefore they do not contribute to the D3-brane charge. This leads to the conclusion that the D3-brane charge of our background is 
\be
\epsilon_{D3}=
-\frac{1}{24}\left(2\chi\left(\faktor{\mathbb{C}^4}{\widetilde{\mathbb{Z}}_4}\right)\right)=-\frac{1}{4}\,,
\ee
which is the D3-brane charge of a $\mathbb{Z}_2$ S-fold.

\begin{center}
{\it $\mathbb{Z}_2$ S-fold  in the presence of an $E_6$ 7-brane}
\end{center}

The fourfold is now described by the following $\mathbb{Z}_6$ quotient of $T^2\times\mathbb{C}^3$: 
\begin{equation}\label{e6z2}
\begin{array}{cccc} 
T^2 & \mathbb{C} & \mathbb{C} & \mathbb{C} \\
\hline 
\gamma_6 & \gamma_6^{-1} & \gamma_6^{3} & \gamma_6^{-3} \\
\end{array}
\end{equation} 
where $\gamma_6^6=1$. We again start from the same kind of  $\mathbb{Z}_6$ quotient for $T^8$: 
\begin{equation}\label{t8orb2}\begin{array}{cccc} 
T^2 & T^2 & T^2 & T^2 \\
\hline 
\gamma_6 & \gamma_6^{-1} & \gamma_6^{3} & \gamma_6^{-3} \\
\end{array}\end{equation}
The Euler characteristic of this fourfold, which we denote as $\faktor{T^8}{\widetilde{\mathbb{Z}}_6}$, was found in \cite{Font:2004et} to be 
$$\chi\left(\faktor{T^8}{\widetilde{\mathbb{Z}}_6}\right)=192.$$ 
In this case there are 16 {fixed point}s: $T^4/\mathbb{Z}_6$ has just one {fixed point} and the remaining $T^4/\mathbb{Z}_2$ has sixteen {fixed point}s. Locally around each point the singularity is a $\mathbb{Z}_6$ orbifold of the form
$$\begin{array}{cccc} 
\mathbb{C} & \mathbb{C} & \mathbb{C} & \mathbb{C} \\
\hline 
\gamma_6 & \gamma_6^{-1} & \gamma_6^{3} & \gamma_6^{-3} \\
\end{array}$$ 
and we will denote this space as $\faktor{\mathbb{C}^4}{\widetilde{\mathbb{Z}}_6}$. Rather as the $\mathbb{Z}_2$-{fixed point}s of the previous case, we do not find any contribution from the $\mathbb{Z}_3$ {fixed point}s since the stabilizer subgroup of these points acts trivially on the last two $T^2$ factors in (\ref{t8orb2}). We have instead a nontrivial contribution from $\mathbb{Z}_2$-{fixed point}s: around such points the singularity is locally $\mathbb{C}^4/\mathbb{Z}_2$, whose Euler characteristic is $3/2$ \cite{Bergman:2009zh}. Next we count the number of such points: on $\faktor{T^2}{\mathbb{Z}_6}$ we have four $\mathbb{Z}_2$-{fixed point}s (including the $\mathbb{Z}_6$-invariant one) and therefore on $\faktor{T^4}{\mathbb{Z}_6}$ there are $4^2-1$ such points (where we have subtracted the $\mathbb{Z}_6$ {fixed point}). We should then divide this number by three to account for the nontrivial action of $\mathbb{Z}_3$ on these points. Finally, we should multiply by 16 (the number of fixed points under the $\mathbb{Z}_2$ action on the last two $T^2$ factors in (\ref{t8orb2})). We therefore conclude that 
\be\ba
&\chi\left(\faktor{T^8}{\widetilde{\mathbb{Z}}_6}\right)=16\chi\left(\faktor{\mathbb{C}^4}{\widetilde{\mathbb{Z}}_6}\right)+\frac{4^2-1}{3}16\frac{3}{2}=192 \cr 
\longrightarrow &\chi\left(\faktor{\mathbb{C}^4}{\widetilde{\mathbb{Z}}_6}\right)=\frac{9}{2}\,.
\ea\ee
In order to compute the D3-brane charge of the geometry (\ref{e6z2}) we have to sum the Euler characteristic of the various fixed points and then multiply by $-\frac{1}{24}$. We have one $\mathbb{Z}_6$-{fixed point}, for which we can use the result we have just derived and we also have the contribution from the $\mathbb{Z}_2$-{fixed point} (there are three of them but they are identified by the $\mathbb{Z}_6$ quotient). We can again neglet $\mathbb{Z}_3$-{fixed point}s since they do not contribute to the D3-brane charge. As a result, we find that the D3-brane charge equals 
\be\ba
\epsilon_{D3}= &
-\frac{1}{24}\left(\chi\left(\faktor{\mathbb{C}^4}{\widetilde{\mathbb{Z}}_6}\right)+\chi\left(\faktor{\mathbb{C}^4}{\mathbb{Z}_2}\right)\right)\cr 
=&-\frac{1}{24}\left(\frac{9}{2}+\frac{3}{2}\right)=-\frac{1}{4} \,.
\ea
\ee
Again we find that the D3-brane charge is equal to that of the underlying S-fold.

\begin{center}
{\it $\mathbb{Z}_3$ S-fold in the presence of a $D_4$ 7-brane}
\end{center}

The $\mathbb{Z}_6$ quotient on $T^2\times\mathbb{C}^3$ is 
\begin{equation}\label{e6z3}
\begin{array}{cccc} 
T^2 & \mathbb{C} & \mathbb{C} & \mathbb{C} \\
\hline 
\gamma_6 & \gamma_6^{-1} & \gamma_6^{2} & \gamma_6^{-2} \\
\end{array}
\end{equation} 
with $\gamma_6^6=1$.
\begin{equation}\label{t8orb3}\begin{array}{cccc} 
T^2 & T^2 & T^2 & T^2 \\
\hline 
\gamma_6 & \gamma_6 & \gamma_6^{2} & \gamma_6^{2} \\
\end{array}
\end{equation}
By a change of coordinate we can rewrite this orbifold action on $T^8$ as  \eqref{e6z3}. 
The Euler characteristic of this fourfold, which we denote as $\faktor{T^8}{\mathbb{Z}'_6}$, was found in \cite{Font:2004et}, and it reads $\chi\left(\faktor{T^8}{\mathbb{Z}'_6}\right)=144$. We repeat exactly the same procedure of the first two cases. The following Euler characteristic has various contributions:
\begin{equation} \label{eq:eulchC4Zp6}
\ba
&\chi\left(\faktor{T^8}{\mathbb{Z}'_6}\right)=9\chi\left(\faktor{\mathbb{C}^4}{\mathbb{Z}'_6}\right)+\frac{3^2-1}{2}9\frac{8}{3}=144 \cr 
\longrightarrow& \chi\left(\faktor{\mathbb{C}^4}{\mathbb{Z}'_6}\right)=\frac{16}{3}\,,
\ea
\end{equation}
where in $T^8$ we have 16 fixed point of the type $\left(\faktor{\mathbb{C}^4}{\mathbb{Z}'_6}\right)$, $(3^3-1)\times 9 \left(\faktor{\mathbb{C}^4}{\mathbb{Z}_3}\right)$, whose Euler characteristic is $\chi\left(\faktor{\mathbb{C}^4}{\mathbb{Z}_3}\right)=\frac{8}{3}$. In addition we have $\faktor{\mathbb{C}^4}{\mathbb{Z}_2}$ fixed points, which can be thought as the 7-brane contributions, and its $\mathbb Z_2$ action is 
\begin{equation}
\begin{array}{cccc} 
T^2 & \mathbb{C} & \mathbb{C} & \mathbb{C} \\
\hline 
\gamma_6^{3} & \gamma_6^{-3} & 1 &1 \\
\end{array}
\end{equation}
On the other hand, these fixed points on $T^8$ contribute as $\mathbb{C}^2/\mathbb{Z}_2\times T^4$, and since $\chi(T^4)=0$, their Euler characteristic also vanishes. Putting everything together the D3-brane charge is 
\begin{equation}
\ba
\epsilon_{D3}= &-\frac{1}{24}\left(\chi\left(\faktor{\mathbb{C}^4}{\mathbb{Z}'_6}\right)+\frac{3-1}{2}\chi\left(\faktor{\mathbb{C}^4}{\mathbb{Z}_3}\right)\right)\cr 
=&-\frac{1}{24}\left(\frac{16}{3}+\frac{8}{3}\right)=-\frac{1}{3}\,.
\ea\end{equation}

For the model with 7-branes {whose backgrounds are not of orbifold type}, i.e. $H_{0,1,2}$, we do not have a direct computation of the D3-brane charge. However, we have observed that the action of the 7-branes leaves always a $T^4$ untouched. {Consequently, it is reasonable to conjecture that like in the case of the orbifold 7-branes, these $H_{0,1,2}$ type 7-brane configurations do not contribute to the D3-brane charge either. This implies that also in those cases, the D3-brane charge is given by the one of the corresponding $\mathcal{N}=3$ S-fold.}

\subsubsection*{Discrete flux}\label{flux} 

We provide here a geometric argument why the D3-brane charge does not depend on the 7-brane even in the presence of $(H_3, F_3)$ discrete flux. For the $\mathcal N=3$ S-fold cases the flux quanta were computed from 
{M-theory on $\faktor{\mathbb C^4}{\mathbb Z_k}$} in \cite{Bergman:2009zh}. The torsion M2-brane charge reads
\begin{equation}
Q_{\rm M2}^{\rm torsion} = -\frac{1}{2} \int_{\mathbb C^4/ \mathbb Z_k} \frac{G_4}{2\pi} \wedge \frac{G_4}{2\pi} =- \frac{1}{2} \int_{S^7/ \mathbb Z_k} \frac{G_4}{2\pi} \wedge\frac{C_3}{2\pi}\,,
\end{equation}
where $G_4=d C_3$ is the 4-form flux of M-theory. In order to compute the integral one needs to construct the explicit 4-dimensional submanifold $\mathcal W$ whose boundary is the torsion cycle $S^3/ \mathbb Z_k$. The IIB 3-brane flux is then computed by reducing the M2-branes on a wrapped circle, and subsequently T-dualizing the system to IIB on a circle prependicular to the IIA O2-plane/D2-brane system \cite{Aharony:2016kai}. The the $G_4$ flux quanta is
\begin{equation}
\int_{\mathcal W} \frac{G_4}{2\pi}= \frac{l}{k} \,,\qquad l \in \mathbb Z\,.
\end{equation}

For this purpose it is useful to parametrize the $S^7$ coordinates as follows:
\begin{eqnarray}
& z_1 = {\rm cos}\left(\frac{\xi}{2}\right)  {\rm cos}\left(\frac{\theta_1}{2}\right) e^{\frac{i(\psi_1+\phi_1)}{2}}\\
& z_2 = {\rm cos}\left(\frac{\xi}{2}\right)  {\rm sin}\left(\frac{\theta_1}{2}\right) e^{\frac{i(\psi_1-\phi_1)}{2}}\\
& z_3 = {\rm sin}\left(\frac{\xi}{2}\right)  {\rm cos}\left(\frac{\theta_2}{2}\right) e^{\frac{i(\psi_2+\phi_2)}{2}}\\
& z_2 = {\rm sin}\left(\frac{\xi}{2}\right)  {\rm sin}\left(\frac{\theta_2}{2}\right) e^{\frac{i(\psi_2-\phi_2)}{2}}
\end{eqnarray}
where $\xi, \theta_1, \theta_2 \in [0,\pi/2]$, $\psi_1, \psi_2 \in [0,2\pi]$, and $\phi_1, \phi_2 \in [0,4\pi]$. The orbifold acts as follows on $S^7$
\begin{equation}
\phi_1 \rightarrow \phi_{1} + \frac{2\pi}{k}, \qquad \phi_2 \rightarrow \phi_{2} + \frac{2\pi}{k\Delta_7}\,.
\end{equation}
and the metric takes the form of two orbifold $S_3$ fibered over an interval $\xi \in [0,\pi/2]$, that is
\be
\ba
ds_{S^7/\mathbb Z_{k\Delta_7} }^2=
 d\xi^2 &+ \frac{ {\rm cos}^2(\xi)}{4} \left(ds_{S^3/\mathbb Z_k}^2\right) \cr 
 &+  \frac{ {\rm sin}^2(\xi)}{4} \left(ds_{S^3/\mathbb Z_{k\Delta_7}}^2\right)\,.
 \ea
\end{equation}
In principle there are two submanifold whose boundary are given by $S^3/\mathbb Z_{k}$ and  $S^3/\mathbb Z_{k\Delta_7}$ at $\xi=0$ and $\xi=\pi/2$ respectively. However, in our background one of the $\mathbb C$ factor of $\mathbb C^2$ is actually compactified resulting in a $T^2$ orbifolded by $\mathbb Z_{k \Delta_7}$. For this reason, one of the boundary  $S^3/\mathbb Z_{k\Delta_7}$ does not exist anymore in our background, and the flux quanta gets only contribution from the torsional $S^3/\mathbb Z_{k}$ exactly as in the $\mathcal N=3$ S-fold case.

All in all we have computed the D3-brane charge with and without flux, and have established that the 7-branes do not contribute explicitly to the value of the D3-brane charge, which are summarized as:
\be\label{charge}\begin{array}{c|c}
k& \text{D3-charge } \epsilon_{D3} \\
\hline 
2 & \frac{1}{4} \\
3 &\frac{1}{3}\\
4 & \frac{3}{8}\\ 
\end{array}\ee
{We will show later that the {a global anomaly} for the combined geometry implies however, that a non-trivial flux has to be turned on for the background to be consistent.}

\subsection{S-folds without Discrete Flux and Anomalies} 

As is known, S-folds that preserve $\mathcal{N}=3$ supersymmetry (at least those with $k<6$), come in two variants, depending on whether we turn on a discrete flux for $H_3$ and $F_3$, or not. The flux contributes to the induced D3-brane charge and we need to take this into account in order to compute the central charges of the resulting $\mathcal{N}=2$ SCFTs. Before discussing this issue, we would like to point out that the field theories we have discussed so far correspond to $\mathcal{N}=2$ S-folds with a nontrivial three-form flux. The easiest way to see this is to notice that the models with enhanced supersymmetry we get upon mass deformation (as described before) correspond to  $\mathcal{N}=3$-preserving S-folds with the discrete flux turned on. 

The natural question regarding our $\mathcal{N}=2$-preserving S-folds is then what happens when we switch off the discrete flux and what are the corresponding 4d theories on the probe D3-brane. Our claim is that this issue does not even arise when we introduce 7-branes because we do not get a consistent background unless we turn on the flux. The argument we would like to provide in support for this conclusion has to do with anomalies and is closely related to the analysis of \cite{Garcia-Etxebarria:2017crf} which we now briefly recall. 

The conclusion of \cite{Garcia-Etxebarria:2017crf} is that 8d SYM with gauge group $G=F_4,\;SO(2N+1)$ is affected by an anomaly measured by $\pi_8(G)$. One way to understand the anomaly is as follows: we put the 8d theory on $S^4\times\mathbb{R}^4$ and we consider an instanton background along $S^4$ for a $SU(2)$ subgroup of the gauge group. When we shrink the sphere we get an effective 4d theory living in $\mathbb{R}^4$ with gauge group $H$, the commutant of $SU(2)$ inside $G$. The fermions charged under $H$ arise from the instanton zero modes of the gaugino. In \cite{Garcia-Etxebarria:2017crf} it was found that for a gauge configuration with instanton number one on $S^4$, the resulting 4d theory is affected by Witten's $\mathbb{Z}_2$ anomaly whenever $G=F_4,\;SO(2N+1)$. The same argument applies in our case as well. This time the 8d gauge fields arising from the 7-branes propagate on $\mathbb{R}^4\times\mathbb{C}^2/\mathbb{Z}_k$ and we can consider an $SU(2)$ instanton on the ALE space instead of $S^4$. 

First of all we should understand what the gauge group is expected to be in the case without three-form flux. As we have seen, since our orbifold involves a $\mathbb{Z}_k$ quotient in the $z_1$ plane transverse to the 7-brane worldvolume, the versal deformations of the Weierstrass are constrained which means that the quotients acts nontrivially on the gauge bundle. Rather as in the case of non-split singularities in F-theory, this action can be identified with a $\mathbb{Z}_k$ outer automorphism of the gauge algebra. Inner automorphisms can in fact be undone with a gauge transformation. 
Let us now focus on the cases $k=2$ and 7-branes of type $E_6$ and $D_4$. In the former case the invariant subgroup under the action of the $\mathbb{Z}_2$ outer automorphism is $F_4$, therefore we conclude that we are dealing with a 8d $F_4$ SYM theory on  $\mathbb{R}^4\times\mathbb{C}^2/\mathbb{Z}_2$\footnote{More precisely we are dealing with an $E_6$ bundle twisted by an outer automorphism around the nontrivial $\mathbb{Z}_2$ cycle. We propose that this does not affect our conclusion since the non-$F_4$ twisted-$E_6$ bundles do not give rise to 4d massless modes.}. In the latter case the invariant subgroup is $SO(7)$, so we expect a 8d $SO(7)$ SYM theory on  $\mathbb{R}^4\times\mathbb{C}^2/\mathbb{Z}_2$. These two theories are anomalous according to \cite{Garcia-Etxebarria:2017crf} and the replacement of $S^4$ with the ALE space $\mathbb{C}^2/\mathbb{Z}_2$ does not change this conclusion. 

In order to see this, let us recall how the fermion zero modes counting works for $SU(2)$ instantons on ALE spaces (we use the conventions of \cite{Bianchi:1996zj}). If we have a $SU(2)$ gauge theory on $\mathbb{C}^2/\mathbb{Z}_2$ with a fermion in the spin $s$ representation, the number of zero modes in an instanton background $I_n^s$ is 
\be\ba
 I_n^s=&\frac{2s(2s+1)(s+1)}{3}n\cr 
 &-\frac{(2s+1)(8s^2+8s+3-3(-1)^{2s})}{24} \,,\ea
 \ee
where $n$ can be identified with the instanton number since the second Chern class of the gauge bundle $\mathcal{E}$ is 
\be c_2(\mathcal{E})=n-\frac{1}{2}\,.\ee
If we consider an $SU(2)$ instanton with $n=1$, $SU(2)$ doublets do not contribute any fermion zero modes, but for $n=2$ we find 
\be\label{counting} I_2^{1/2}=1;\quad I_2^1=6;\quad I_2^{3/2}=14\,.\ee 
Using such an instanton background we can easily recover the anomaly discussed in \cite{Garcia-Etxebarria:2017crf}, therefore providing evidence that our claim is indeed correct: we have to introduce extra ingredients (i.e. the flux) in order to get a consistent background. 

Let us start by discussing the case $G=SO(2N+1)$ with $N>2$, which in particular applies to the 7-brane of type $D_4$ combined with a $k=2$ S-fold. We consider the decomposition 
\be\label{decso} SO(2N+1)\supset SU(2)\times SU(2)'\times SO(2N-3)\,,\ee 
and turn on the instanton for the first $SU(2)$ factor. We therefore have $H=SU(2)'\times SO(2N-3)$ and the adjoint of $G$ decomposes under \eqref{decso} as 
\be\ba
{\bf (2N+1)N}\rightarrow& ({\bf 2}, {\bf 2}, {\bf 2N-3})\oplus 
({\bf 3}, {\bf 1}, {\bf 1}) \cr 
&\oplus ({\bf 1}, {\bf 3}, {\bf 1})\oplus ({\bf 1}, {\bf 1}, {\bf (2N-3)(N-2)}) \,.\ea
\ee
 Using \eqref{counting} we find that the resulting 4d theory has fermions transforming in the representation 
\be ({\bf 2}, {\bf 2N-3})+\text{{singlets under $H$}}\,,\ee 
and this theory is anomalous because it contains an odd number ($2N-3$) of $SU(2)'$ doublets. 

The argument for $F_4$, which is relevant for a 7-brane of type $E_6$ combined with a $k=2$ S-fold, follows from the previous one. It is known that $F_4\supset SO(9)$ and the adjoint representation decomposes into the adjoint and spinor of $SO(9)$ 
\be {\bf 52}\rightarrow {\bf 36}+{\bf 16}\ee 
As we have seen before, the adjoint of $SO(9)$ leads to an anomalous theory in 4d and therefore, unless the contribution from the 16-dimensional spinor representation cancels the anomaly, the resulting theory will be anomalous. As before, we consider the subgroup $SU(2)\times SU(2)'\times SO(5)$, under which the spinor of $SO(9)$ decomposes as 
\be {\bf 16}\rightarrow  ({\bf 2}, {\bf 1}, {\bf 4})+ ({\bf 1}, {\bf 2}, {\bf 4})\,.\ee 
From this decomposition we see that, if we turn on the instanton for $SU(2)$, the spinor of $SO(9)$ does not contribute to the $SU(2)'$ anomaly and therefore we conclude that the $F_4$ theory as well is anomalous.

{This confirms our expectation that the theories without flux will be inconsistent and we therefore have to turn on flux in the presence of the 7-branes with S-folds.}
{We should emphasize however that the direct check of this is only performed for the case $k=2$ and 7-branes of type $E_6$ and $D_4$. In the other cases it would be very interesting to find a similarly compelling argument. }

\subsection{S-folds and Discrete Flux} 

Although we do not have a detailed understanding of the mechanism which resolves the anomaly issue we have just discussed, once the flux is included, let us attempt to give a qualitative explanation. As is well known, one consequence of the inclusion of torsional flux is a change in the topology of the gauge bundle due to the Freed-Witten (FW) anomaly \cite{Freed:1999vc} (actually its extension proposed by Kapustin \cite{Kapustin:1999di}). In our case the manifold wrapped by the 7-branes has vanishing $W_3$ class, as was pointed out for $k=2$ in \cite{Witten:1998xy}, therefore the FW equation relates the cohomology class of the flux to the topological class of the gauge bundle, which is measured by $H^2(M,\pi_1(G))$ (where $M$ is the base, in our case a ALE space). For this quantity to be nontrivial, the gauge group $G$ cannot be a simply-connected Lie group, but rather a quotient thereof by (a subgroup of) the center. This suggests that in the presence of discrete flux we are dealing with the so-called gauge bundles without vector structure, which is not surprising after all since such bundles are a key ingredient of the construction of rank one theories from compactification of 6d SCFTs \cite{Ohmori:2018ona}. 

The quotient along the $z_1$-direction is now accompanied by a $\mathbb{Z}_k$ automorphism of the gauge bundle, which however does not necessarily coincide with the automorphism group of the underlying Lie algebra. In summary, the $\mathbb{Z}_k$ projection we have to consider once the flux is turned on may differ from the usual outer automorphism we have discussed before and therefore the invariant subgroup, which is the crucial piece of information for the anomaly computation presented above, can change. Let us discuss in more detail this point for the $k=2$ cases analyzed before. In the case of $E_6$, besides the outer automorphism with invariant subgroup $F_4$, there is a $\mathbb{Z}_2$ projection whose invariant subgroup is $Sp(4)$. If this is the relevant $\mathbb{Z}_2$ we have to consider, then we end up with a different twisted-$E_6$ bundle in 8d. The fact that the invariant subgroup is $Sp(4)$ suggests that the anomaly should not be an issue anymore, since after all 8d SYM with symplectic gauge group arises in string theory by combining D7-branes with a $O7^+$ orientifold plane. Analogously, the invariant subgroup of $SO(8)$ under the $\mathbb{Z}_2$ outer automorphism is $SO(7)$ which is anomalous, but there is also another $\mathbb{Z}_2$ projection whose invariant subgroup is $Sp(2)\times SU(2)$ and again the anomaly is not an issue in this case. 

Based on this considerations, we expect the global symmetry $G_F$ of the 4d theory living on the worldvolume of the D3-brane to contain as a subgroup:
{ 
\be\label{globalguess}\begin{array}{|c|c|}
\hline
\text{Theory} \langle G,  \mathbb{Z}_k\rangle & G\subset G_F \\
\hline 
\langle E_6, \bbZ_2\rangle & Sp(4){\color{blue}\times SU(2)} \\ 
\hline 
 \langle D_4, \bbZ_2\rangle & Sp(2)\times SU(2){\color{blue}\times SU(2)} \\
\hline 
 \langle H_2, \bbZ_2\rangle & SU(2)\times U(1){\color{blue}\times SU(2)} \\
\hline 
 \langle D_4, \bbZ_3\rangle & SU(3){\color{blue}\times U(1)} \\ 
\hline 
 \langle H_1, \bbZ_3\rangle & U(1){\color{blue}\times U(1)} \\
\hline 
 \langle H_2, \bbZ_4\rangle & SU(2){\color{blue}\times U(1)} \\ 
\hline
\end{array}\ee
}
 In the table we have included in blue the global symmetry coming from the isometries of the background and in black (a choice for) the invariant subgroup under the $\mathbb{Z}_k$ quotient. Notice that in all cases we get subgroups of the actual global symmetry of the corresponding SCFT, which is known. We do not capture the full global symmetry of the theory because the subgroup arising from 7-branes and the subgroup coming from isometries are not on equal footing in our setup, whereas in the 4d theory they fit inside a larger group $G_F$ which we do not see.  As we will see momentarily, there is a good reason for this. The enhancement of global symmetry is an "accident'' of the rank one case and does not occur for higher rank models. 
{Such accidental low rank enhancements are quite common, also in other dimensions.} Since in our setup increasing the rank just corresponds to introducing more D3-branes, we can at best see the global symmetry which is common to arbitrary rank theories. We will provide below one example in which the rank two version of the theory has exactly the global symmetry $G$ appearing in \eqref{globalguess}.


\section{Central Charges, Higher Rank and Holography}
\label{sec:CC}

In order to compute the {central charges of the SCFTs}, we implement the method introduced in \cite{Aharony:2007dj}. Having {computed the } D3-brane charges in \eqref{charge}, we can now study the near-horizon geometry of the F-theory background. We recall that being defined in F-theory the background does not need to have a small string coupling $g_s$, {and such F-theoretic holography setups are by now well-studied (see e.g. \cite{Fayyazuddin:1998fb, Couzens:2017way, Couzens:2017nnr}).} The strategy would be to extract the contribution at orders $O(N^2), O(N)$. The order $O(0)$ is somewhat trickier to estimate and depends on the number of free hypermultiplet at any point of the Coulomb branch.

\subsection{Central Charges}

As already anticipated, we can view our $\mathcal{N}=2$ S-folds as a two-fold action on the space resulting of an $\mathcal N=3$ S-fold on top of a 7-brane background. This point of view is useful to define the holographic limit.
Let us first start with the 7-brane background in flat-space. The geometry reads \cite{Fayyazuddin:1998fb}
\begin{equation}
ds^2= ds^2_{{\rm mink}_8} + {\frac{1}{\Delta_7^2}}|z^{-\frac{n_7}{12}}dz|^2, \qquad n_7=12 \frac{(\Delta_7-1)}{\Delta_7}\,,
\end{equation}
where $z=x^8 + ix^9$. The transverse space to the 7-brane corresponds to the orbifolds $\mathbb C/ \mathbb Z_{\Delta_7}$ with $\Delta_7=1, \frac{6}{5}, \frac{4}{3}, \frac{3}{2},2 ,3,4,6$ respectively\footnote{In order to treat all these cases in a unified fashion, the notation is slightly simplified: For instance, when $\Delta_7$ is not integer (for $H_0,H_1,H_2$), the backgrounds are not really global orbifolds. On the other hand, the important data for the central charge computation is the deficit angle $\frac{2\pi}{\Delta_7}$, which is well defined for any $\Delta_7$.}. The coordinate of $\mathbb C$ is $u$, and we have that 
\begin{equation}
z=u^{\Delta_7}\,.
\end{equation} 
The orbifold action is $u\rightarrow e^{2\pi i/\Delta_7} u$, which implies that $|z^{-n_7/12}dz|^2={ \Delta_7^2} du^2$.

{In addition to this 7-brane background we introduce the S-fold action}. The S-fold acts on $\mathbb C/ \mathbb Z_{\Delta_7}$ and on a $\mathbb{C}^2$ inside the 8-dimensional space wrapped by the 7-branes. The $\mathbb{C}^2$ is parametrized by the coordinates $v=x^4 + i x^5$ and $\tilde v=x^6 + i x^7$. The action can be summarized as follows
\begin{align}
& v \rightarrow e^{\frac{2\pi i}{ k}} v\\
& \tilde v \rightarrow e^{-\frac{2\pi i}{ k}} \tilde v\\
& z \rightarrow e^{\frac{2\pi i}{ k}} z.
\end{align}
The last orbifold action on the coordinate $u$ is given by
\begin{equation} \label{eq:kdeltaid}
u \rightarrow e^{\frac{2\pi i}{ k \Delta_7}} u \, .
\end{equation}
When $N$ D3-branes probe this geometry we can take a near-horizon limit where we first implement the following spherical change of coordinate
\begin{align}
&{u=r {\rm cos}(\phi) e^{ i \theta}}\\  
&v=r \,{\rm sin}(\phi) {\rm cos}(\beta) e^{i \omega}\\
&v=r \,{\rm sin}(\phi) {\rm sin}(\beta) e^{ i \tilde \omega}\,,
\end{align}
where $\theta, \omega, \tilde \omega \in [0,2\pi]$ and $\phi, \beta \in [0,\pi/2]$. The coordinates $\beta, \omega, \tilde \omega$ parametrize an $S^3$, and the $\mathbb{Z}_k$ symmetry acts freely on it. 

The near-horizon geometry is 
\begin{widetext}
\begin{equation}
ds^2=\alpha' \left( \frac{\alpha'}{R^2} \rho^2 ds^2_{{\rm mink}_4} + \frac{R^2}{\alpha'} \frac{d \rho^2}{\rho^2} + \frac{R^2}{\alpha'} \left( d\phi^2 +\frac{ {\rm cos}^2(\phi)}{k^2\Delta_7^2} d\theta^2 + {\rm sin}^2(\phi) d\Omega_{S^3/\mathbb{Z}_{ k}} \right) \right)\,,
\end{equation}
\end{widetext}
where $R$ is the radius of AdS$_5$ and $M_5=(S^5/\mathbb Z_{\Delta_7})/\mathbb Z_{ k}$, and the second orbifold is a free action on $S^5$. The flux quantization for $F_5$ reads
\begin{equation}
\int_{M_5} F_5= N+ \epsilon_{D3}\,.
\end{equation}
where $ \epsilon_{D3}$ is the charge of the S-fold \eqref{charge}. It follows that the radius $R$ is related to squared YM-coupling, the number of D3-branes, $N$, and the S-fold charge
\begin{equation} \label{eq:AdSR}
 \frac{R^2}{\alpha'} = \sqrt{4\pi g^2_{YM} (N+ \epsilon_{D3})}.
\end{equation}
This completes what we need to know about the holographic dual, and we are ready to compute the central charges. 

\paragraph{Bulk contribution:} in order to evaluate the contribution from the bulk, which contains the leading order behavior of the central charges, $O(N^2)$, we follow the procedure of \cite{Henningson:1998gx}. {The bulk action goes as follows,
\begin{equation} \label{eq:bulkact}
S_{\text{bulk}}\sim \frac{R^8 {\rm Vol}(M_5)}{G_N^{(10)}  {\rm Vol}(M_5)^2 }
\end{equation}
where the ten-dimensional Newton constant is $G_N^{(10)}\sim g_s^2 \alpha'^4$ and $g_s\sim g_{YM}^2$.
We then plug into \eqref{eq:bulkact} the value of $R$ in \eqref{eq:AdSR}, similarly to \cite{Aharony:2007dj} we get}
\begin{equation}
a(\text{bulk})=c(\text{bulk})=\frac{(N+ \epsilon_{D3})^2}{4 {\rm Vol}(M_5)} \,, 
\end{equation}
where we normalized the volume of $M_5$ such that $\pi$ does not appear. By substituting the integrated volume form on $M_5$ we get
\begin{equation} \label{eq:quadordcc}
a(\text{bulk})=c(\text{bulk})= k \Delta_7 \frac{(N+ \epsilon_{D3})^2}{4}\,.
\end{equation}
This contribution is similar to the central for the $\mathcal N=3$ theories computed in \cite{Aharony:2016kai}, indeed it reduces to it when $\Delta_7=1$, where in this case the $\epsilon_{D3}^2$ contribution is canceled by the $O(0)$ order, see appendix \ref{app:CSP}.

\paragraph{Subleading order behaviour:} to compute this contribution we again implement the procedure highlighted in \cite{Aharony:2007dj}. The expression we need to evaluate is the Chern-Simons term of the 7-brane worldvolume action 
\begin{equation}
S_{\rm CS_7}\sim A n_7 \int C_4 \wedge \left( {\rm tr}( R_T \wedge R_T)- {\rm tr}( R_N \wedge R_N) \right)
\end{equation}
where $R_T$ is the curvature of the tangent bundle to the 7-branes, whereas $R_N$ is the curvature of the normal bundle. In AdS$_5$ they will give rise to terms proportional to $U(1)_R SU(2)_{R,L}$ and  $U(1)_R^3$, $p_1(TM_4)U(1)_R$ respectively. 

On dimensional ground we can estimate that the 7-brane CS-action is proportional to 
\begin{equation}
S_{\rm CS_7} { \sim \frac{R^4}{g_s \alpha'^2}} \frac{{\rm Vol}(S^3/\mathbb Z_{ k})}{{\rm Vol}(M_5)}.
\end{equation}
We substitute the integrated volumes and we normalize the contribution like \cite{Aharony:2007dj}, such that for $ k=1$ it reproduces their result. The  linear contribution of the central charges is given by
\be \label{eq:Nord}
\ba
a(\text{7-brane})&=\frac{ k \Delta_7 n_7 (N+ \epsilon_{D3})}{24  k}= \frac{  (N+ \epsilon_{D3}) (\Delta_7 -1)}{2 },\\
c(\text{7-brane})&=\frac{ k \Delta_7 n_7  (N+ \epsilon_{D3})}{16  k}=\frac{ 3 (N+  \epsilon_{D3}) (\Delta_7 -1)}{4 }.
\ea
\ee

\paragraph{Lowest order behaviour, $O(N^0)$:} we need to evaluate the smallest contribution which comes at order zero in $N$. This contribution comes from massless hypermultiplets which are present at any point of the coulomb branch, which are neutral under the gauge symmetry and span what in \cite{Argyres:2016yzz} is called enhanced Coulomb branch. One of these hypermultiplet is associated to closed string modes of IIB supergravity localizing on the D3-brane. In appendix \ref{app:CSP}, we look at the closed string spectrum as in \cite{Bilal:1999ph, Fayyazuddin:1998fb, Aharony:2007dj}. The outcome of this analysis is
\begin{itemize}
\item for $k=1$ this hypermultiplet is always projected out similarly to the $\mathcal N=4$ case. 
\item When the S-fold group is non-trivial, this  hypermultiplet survives.
\item When $k>1$ and $\Delta_7 \neq 1$, there are additional massless hypermultiplets on top of the one corresponding to the IIB supergravity closed string mode localizing on the D3-brane. These are coming from the effective theory engineered by the F-theory background. In particular, as explained in section \ref{sec:ECB}, see also \cite{Argyres:2016yzz}, when the D3-brane is away from the S-fold singularity, the theory is described by a free vector and $h$ massless free hypers, which are counted by
\begin{equation}
h=k(\Delta_7-1) +1.
\end{equation} 
\end{itemize} 

We now write the full formula for the central charges, where we have removed the $O(N^0)$ terms coming from $\epsilon_{D3}$ and $\epsilon_{D3}^2$ contribution in \eqref{eq:Nord} and \eqref{eq:quadordcc} in order to properly count the massless hypermultiplets,
\be\label{centralc}
\ba
&a = \frac{  k \Delta_7 N^2  }{4} +\frac{ N(k \Delta_7 +\Delta_7 -2)}{4 } +   \frac{ k (\Delta_7-1)}{24}\\
&c=  \frac{ k \Delta_7 N^2  }{4} +\frac{ N (k \Delta_7 +2 \Delta_7 -3)}{4 } +   \frac{ k (\Delta_7-1)}{12} 
\ea
\ee
We need to consider that, however, the last term for $k=1$ is actually not there, since in that case for $N=0$ and $k=1$ there is no S-fold, and the theory is not 4-dimensional without a D3-brane. Notice that for $\Delta_7=1$ the formula reproduce the central charges for the $\mathcal{N}=3$ theories. 

Remarkably, for $N=1$ the formula \eqref{centralc} perfectly reproduces the central charges of known rank one theories according to our identification \eqref{ident4d}. Notice that the inclusion of the $O(N^0)$ terms in \eqref{centralc} (i.e. the contribution from the free hypers) is crucial for reproducing the central charges reported in Table \ref{tab1}.
 
\subsection{Coulomb Branch Operators and Higher Rank Theories}
The Coulomb branch of higher rank $\mathcal{N}=2$ S-fold theories is parametrized by the positions of $N$ D3-branes along the direction orthogonal to the 7-branes, that is $\mathbb{C}/\mathbb{Z}_{k\Delta_7}$. The positions are denoted by $u_i$, where $i=1,\ldots,N$. The Coulomb branch moduli space is symmetric with respect to the exchange of D3-brane positions, $u_i \leftrightarrow u_j$. Moreover we have the $\mathbb{Z}_{k\Delta_7}$ symmetry coming from the S-fold and 7-branes acting on $\mathbb{C}$, and therefore on the $u_i$, 
\begin{equation}
u_i \rightarrow e^{\frac{2 \pi i}{k\Delta_7}}u_i,
\end{equation}
This resembles the background identification on the $u$ coordinate in \eqref{eq:kdeltaid}. The gauge invariant operators in the Coulomb branch are then identified with the polynomials in $u_i$ which respect the symmetries of the background, as well as the one exchanging the D3-branes. They correspond to the following symmetric polynomials,
\begin{equation}
\sum_{i=1}^{N} u_i^{j k\Delta_7}, \qquad j=1,\ldots, N.
\end{equation}
The scaling dimensions of these polynomial are measured by their exponents, and they read, 
\begin{equation}
k \Delta_7, 2 k \Delta_7, \ldots, N k \Delta_7.
\end{equation}
As a further highly nontrivial consistency check of our construction, we notice that at least one of these theories has already been analyzed in the context of class $\mathcal{S}$ theories: The rank two version of the $[II^*,C_5]$ theory is a trinion of the twisted $E_6$ theory \cite{Chacaltana:2015bna}, exactly with central charges \eqref{centralc}. Notice that the global symmetry reported in  \cite{Chacaltana:2015bna} is $Sp(4)\times SU(2)$, which is precisely the symmetry found in  \eqref{globalguess}, without any further enhancement. 

We can also easily predict the properties of Higgs Branch flows and the dimension of the Enhanced Coulomb Branch. As we move the stack of D3-branes away from the singular point along the directions $z_{2,3}$, the information about the orbifold is lost and the probe branes only see the 7-branes. The effective low energy theory at these special points of the Higgs Branch is therefore identified with the rank $N$ version of the $H_1$, $H_2$, $D_4$ or $E_6$ SCFTs, depending on the case, plus a collection of $k(\Delta_7-1)+1$ free hypermultiplets. We therefore conclude that the dimension of the Enhanced Coulomb Branch is 
\be
h = k(\Delta_7-1)+N \,.
\ee

\begin{acknowledgments}
We thank A. Braun,  M. Martone, C. Meneghelli, J. F. Morales, W. Peelaers, Y.-N. Wang and T. Weigand for discussions and Y. Tachikawa for detailed comments on the draft. This work is
supported by the ERC Consolidator Grant number 682608 Higgs bundles: Supersymmetric
Gauge Theories and Geometry (HIGGSBNDL).

\end{acknowledgments}


\appendix

\section{Alternative Computation of Orbifold Euler Characteristics\label{app:D3alt}}
The D3-brane charge computation in presence of an orbifold relies on the Euler characteristic of the fixed points, \cite{Bergman:2009zh}. In section \ref{noflux} we have computed the Euler characteristic of the fixed points by using results of perturbative string theory on orbifolded tori \cite{Font:2004et}. As pointed out in \cite{Bergman:2009zh}, one also estimate the Euler characteristic of the fixed points by removing boundary contributions. For instance, in case of the following actions,
\begin{equation}
\begin{array}{cccc} 
\mathbb C & \mathbb{C} & \mathbb{C} & \mathbb{C} \\
\hline 
\gamma_k & \gamma_k^{-1} & \gamma_k & \gamma_k^{-1}\\
\end{array}
\end{equation} 
for $\gamma_k = e^{\frac{2\pi i}{k}} \in \mathbb Z_k$ and $k=2,3,4,6$, we have that 
\begin{equation}\ba
&\chi_{\rm fixed\; pnts}\left( \faktor{\mathbb C^4}{\mathbb Z_k}\right)\cr 
&=\chi_{\rm bulk}\left( \faktor{\mathbb C^4}{\mathbb Z_k}\right)=\chi \left( \faktor{\mathbb C^4}{\mathbb Z_k}\right)- \chi_{\partial}\left( \faktor{\mathbb C^4}{\mathbb Z_k}\right)\cr 
&= k -\frac{1}{k}\,,
\ea
\end{equation}
where the contribution from bulk and boundary $\partial$ are
\begin{eqnarray}
& \chi_{\rm bulk}\left(\mathcal M \right) = \int_M  e(T\mathcal M) = k \\
&  \chi_{\partial}\left(\mathcal M \right) = \int_{\partial M}  {\rm CS}_7(\omega)=\frac{1}{k} \,,
\end{eqnarray}
where the last equalities are valid because $\mathcal{M}= \faktor{\mathbb C^4}{\mathbb Z_k}$ \cite{Mohri:1997ef}, and $\partial \mathcal{M}= \faktor{ S^7}{\mathbb Z_k}$ with a $\mathbb Z_k$ free action, moreover  ${\rm CS}_7(\omega)$ is the Chern-Simons 7-form for the spin connection $\omega$.

We want now to implement this method for more complicated orbifold actions, that are
\begin{equation} \label{eq:orbact}
\begin{array}{ccccc} 
&\mathbb C & \mathbb{C} & \mathbb{C} & \mathbb{C} \\
\cline{2-5}
\widetilde \gamma_4= &\gamma_4 & \gamma_4^{-1} & \gamma_4^2 & \gamma_4^{-2}\\
\cline{2-5}
\widetilde\gamma_6= &\gamma_6 & \gamma_6^{-1} & \gamma_6^3 & \gamma_6^{-3}\\
\cline{2-5}
\widetilde\gamma_6'= &\gamma_6 & \gamma_6^{-1} & \gamma_6^2 & \gamma_6^{-2}\\
\end{array}
\end{equation} 
In particular the some of the orbifold actions will have the following set of fixed loci:
\begin{equation} \label{eq:fxdpntloci}
\begin{array}{cccc} 
\widetilde\gamma_4^2 & \quad & \rightarrow & {\rm fixed\; locus:}\;\faktor{\mathbb C^2}{\mathbb Z_2}\\
\{ \widetilde\gamma_6^{2}, \, \widetilde\gamma_6^{4} \}   & \quad & \rightarrow & {\rm fixed\; locus:}\; \faktor{\mathbb C^2}{\mathbb Z_2}\\
 \widetilde\gamma_6^{'3} & \quad & \rightarrow &{\rm fixed\; locus:}\; \faktor{\mathbb C^2}{\mathbb Z_3}\\
\end{array}
\end{equation}
The Euler characteristic receives contribution from these fixed loci as well. In fact we have that 
\begin{equation} \label{eq:bulkbundformfix}\ba
&\chi_{\rm bulk}\left( \faktor{\mathbb C^4}{ \widetilde{\mathbb  Z}_k}\right) \cr 
&=\chi \left( \faktor{\mathbb C^4}{ \widetilde{\mathbb  Z}_k}\right)- \chi_{\text{free}-\partial}\left( \faktor{\mathbb C^4}{ \widetilde{\mathbb  Z}_k}\right) - \chi_{\text{fixed}-\partial }\left( \faktor{\mathbb C^4}{ \widetilde{\mathbb  Z}_k}\right) 
\ea\end{equation}
where $\partial$ stands for boundary, and where in this case we have extra piece,  due to
$ \chi_{\text{fixed}-\partial}\left( \faktor{\mathbb C^4}{ \widetilde{\mathbb  Z}_k}\right)$ listed in \eqref{eq:fxdpntloci}. In other words this is actually the contribution of the fixed loci at the boundary, which are
\begin{equation} \label{eq:bndryfxdpntloci}
\begin{array}{cccc} 
\widetilde\gamma_4^2 & \quad & \rightarrow & {\rm boundary\; fixed\; locus:}\;\faktor{ S^3}{\mathbb Z_2}\\
\{ \widetilde\gamma_6^{2}, \, \widetilde\gamma_6^{4} \}   & \quad & \rightarrow & {\rm  boundary\; fixed\; locus:}\; \faktor{ S^3}{\mathbb Z_2}\\
 \widetilde\gamma_6^{'3} & \quad & \rightarrow &{\rm  boundary\; fixed\; locus:}\; \faktor{\mathbb S^3}{\mathbb{Z}_3}\\
\end{array}
\end{equation}
where all action on $S^3$ are free, and its Euler characteristic reads
\begin{eqnarray}
 \chi_{\partial}\left( \faktor{\mathbb C^2}{\mathbb Z_2} \right) &= \int_{\partial \left(\faktor{\mathbb C^2}{\mathbb Z_2} \right)}  {\rm CS}_3(\omega) \cr 
&= \int_{\faktor{\mathbb S^3}{\mathbb Z_2} }  {\rm CS}_3(\omega)=\frac{1}{2} \\
 \chi_{\partial}\left( \faktor{\mathbb C^2}{\mathbb Z_3} \right)& = \int_{\partial \left(\faktor{\mathbb C^2}{\mathbb Z_3} \right)}  {\rm CS}_3(\omega)\cr 
 &=  \int_{\faktor{\mathbb S^3}{\mathbb Z_3} }  {\rm CS}_3(\omega)=\frac{1}{3}  \,.
\end{eqnarray}
The final result is
\be
\ba
 \chi_{\rm bulk}\left( \faktor{\mathbb C^4}{ \widetilde{\mathbb  Z}_k}\right)&= 4 - \frac{1}{2} - \frac{1}{2}  = 3 \, ,\\
 \chi_{\rm bulk}\left( \faktor{\mathbb C^4}{ \widetilde{\mathbb  Z}_k}\right)&= 6 - \frac{1}{2} - 2 \times  \frac{1}{2}  = \frac{9}{2} \, , \\
 \chi_{\rm bulk}\left( \faktor{\mathbb C^4}{ \widetilde{\mathbb  Z}_k}\right)&= 6 - \frac{1}{3} -  \times  \frac{1}{3}  = \frac{16}{3} \, ,
\ea
\ee
which matches with the characters computed in section  \ref{noflux}, and the rest of the computation for the D3-brane charge is completely analogous. 

\section{Closed String Spectrum on AdS$_5 \times M_5$} \label{app:CSP}

As we anticipated, one way to detect the contribution at zero'th order in $N$ is to consider the closed string spectrum. For $\mathcal N=4$ it is enough to look at the fermionic spectrum in AdS$_5 \times S^5$, which contains the following tower of states denoted by $\lambda$ \cite{Kim:1985ez,Gunaydin:1985cu},
\begin{equation} \label{eq:fermspect}\begin{array}{c|c|c}
  (j_1,j_2)&  m & SU(4)_R \\
\hline
  \left(\frac{1}{2},0 \right) & - \left(k+\frac{7}{2} \right) \;\; k\geq 0& 4^*, 20^*, \ldots \\
\hline
  \left(\frac{1}{2},0 \right) & - \left(k-\frac{1}{2} \right) \;\; k\geq 1& 20^*, \ldots 
\end{array}
\end{equation}
where $(j_1,j_2)$ are the $SU(2) \times SU(2)$ spins of $SO(1,3)$ Lorentz, whereas $m^2=\Delta(\Delta-4)$ with $\Delta$ the conformal dimension of dual operators. This tower is the only fermionic tower with a discrepancy of the spectrum. Indeed we can observe that in the second line \eqref{eq:fermspect} there is a missing $4$ representation of the $\mathcal N=4$ $SU(4)$ R-symmetry. In \cite{Bilal:1999ph} it was shown that this discrepancy leads to the following $ O(1)$ contribution to the central charge
\begin{equation}
a(1)=c(1)=-\frac{1}{4}\,.
\end{equation}
This was by computing the chiral anomaly induced by these modes. This contribution corresponds to the decoupling of the center of mass $U(1)$ of the D3-brane stack. 

Our geometry can be seen as a combination of two kind of quotients of $S^5$. Let us separately analyze the effect of these two quotients on the tower \eqref{eq:fermspect}. The other fermionic states will not contribuite since there is no discrepancy in the R-symmetry representations. 

\underline{\it Free S-fold quotient $\mathcal N=3$:} 
In this case the R-symmetry $SU(4)_R$ becomes $SU(3)_R \times U(1)_P$, \cite{Garcia-Etxebarria:2015wns}. We have then the following branching rule for the fermionic tower $\lambda$
\begin{equation}
\mathbf{4} \rightarrow \mathbf 1_{-3}\oplus \mathbf 3_{1}.
\end{equation}
This modes acquire a phase under the free orbifold action of $\mathbb Z_{ k}$ on $S^5$, which is given by
\begin{equation}
e^{\pi i \frac{q_P+1}{ k}}
\end{equation}
where the $+1$ shift is due to the fact that we are considering fermionic modes. As we can see since $q_P=1,-3$, these phases are non-invariant and, therefore, these states are all projected out from the spectrum. This leads to no chiral discrepancy, and no order zero contribution to the central charges of $ k >2$. For this reason, we need to subtract the following contribution from \eqref{eq:quadordcc}, $k  \epsilon_{D3}^2/4$, and the number of gauge neutral massless hypermultiplets in the CB is $h=1$

Moreover, since the action is free on $S^5$ there are no twisted states to analyze. 

\underline{\it 7-brane type quotient $\mathcal{N}=2$:} In this case the $SU(4)_R$ breaks into $SU(2)_L \times SU(2)_R \times U(1)_R$. Consequently we have the following branching rule
\begin{equation}
\mathbf{4} \rightarrow \mathbf{(2,1)}_{-1}\oplus \mathbf{(1,2)}_{1}.
\end{equation}
Again under the orbifold action these modes acquire a phase 
\begin{equation}
e^{\pi i \frac{q_R+1}{\Delta_7}}
\end{equation}
where $q_R$ is the charge of this fermionic tower under $U(1)_R$. As we can see the second state is projected out and does not contribute to the discrepancy. $\mathbf{(2,1)}_{-1}$ is associated to the decoupling of the free hyper which transforms as the fundamental of $SU(2)_L$. This contributes $a(1)=-\frac{1}{24}$ and $a(1)=-\frac{1}{12}$ \cite{Aharony:2007dj}. In this case the number of gauge neutral massless hypermultiplets in the CB is $h=0$.

\underline{\it {S-fold plus 7-branes}, $\mathcal{N}=2$:} Let us take the $SU(4)_R \rightarrow SU(2)_L \times SU(2)_R \times U(1)_R$ R-symmetry fermionic tower 
\begin{equation}
\mathbf{4} \rightarrow \mathbf{(2,1)}_{-1}\oplus \mathbf{(1,2)}_{1}.
\end{equation}
The S-fold breaks this symmetry into
$ SU(2)_R \times U(1)_R \times U(1)_L$, and we have 
\begin{equation}
\mathbf{4} \rightarrow \mathbf{(2,1)}_{-1}\oplus \mathbf{(1,2)}_{1}\rightarrow  \mathbf{1}_{(-1,-1)}\oplus \mathbf{1}_{(-1,1)}\oplus \mathbf{2}_{(1,0)}.
\end{equation}
Under the orbifold action these modes acquire a phase 
\begin{equation}
e^{\pi i \frac{q_R+q_L+1}{k}}.
\end{equation}
By plugging in the charges we get that none of these states survives. As in the $\mathcal N=3$ case there is no order zero contribution from the closed string spectrum, such that the gauge neutral massless hypermultiplets in the CB is (at least 1) $h\geq1$. {As already anticipated, we followed the procedure of \cite{Bilal:1999ph}, which gives the expected result in terms of the supergravity spectrum on AdS$_5 \times M_5$, dual to $\mathcal{N}=2$ theories coming from D3-branes probing the S-fold plus 7-branes background. In \cite{Liu:2010gz,Ardehali:2013xya}, $(c-a)$ was computed from the spectrum of supergravity on AdS$_5 \times M_5$, where $M_5$ is an orbifold of $S^5$. It was shown that for dual $\mathcal{N}=1$ theories this contribution can be rather non-trivial. It would be interesting to apply this approach to our backgrounds and to fully reproduce  $(c-a)$, in particular the 7-brane contribution as well as the zeroth order one. In addition, \cite{Ardehali:2013xya} analyzes the twisted modes which are present when the orbifold produces an $S^1$ fixed-locus on $S^5$. In our case the geometry of the fixed-loci is different, i.e. AdS$_5 \times S^3/\mathbb{Z}_k$, and the twisted modes come from reduction of the 8d theory on the 7-branes reduced on $S^3/\mathbb{Z}_k$. We believe that these twisted modes reproduce the contribution of the rest of the free hypermultiplets, counted by $k(\Delta_7 -1)$. It would be interesting to explore this further. }

%

\end{document}